\documentclass[english]{IEEEtran}
\usepackage[LGR,T1]{fontenc}
\usepackage[active]{srcltx}
\usepackage{color}
\usepackage{array}
\usepackage{verbatim}
\usepackage{units}
\usepackage{multirow}
\usepackage{amsmath}
\usepackage{graphicx}

\usepackage[utf8]{inputenc}
\usepackage[english]{babel}
\usepackage{fancyhdr}
\makeatletter

\DeclareRobustCommand{\greektext}{%
  \fontencoding{LGR}\selectfont\def\encodingdefault{LGR}}
\DeclareRobustCommand{\textgreek}[1]{\leavevmode{\greektext #1}}
\ProvideTextCommand{\~}{LGR}[1]{\char126#1}

\providecommand{\tabularnewline}{\\}

\usepackage{graphicx}
\usepackage{caption}
\usepackage{cite}

\makeatother

\usepackage{babel}
\begin{document}
\title{Numerical and Experimental Characterization of LoRa-Based Helmet-to-Unmanned Aerial Vehicle Links on Flat Lands\\ \LARGE A Numerical-Statistical Approach to Link Modeling.}
\author{\IEEEauthorblockN{G. M. Bianco, A. Mejia-Aguilar, G. Marrocco}}
\maketitle
\begin{abstract}
The use of the LoRa communication protocol in a new generation of
transceivers is attractive for search and rescue (SaR) procedures
because they can operate in harsh environments covering vast areas
while maintaining a low power consumption. The possibility of wearing
helmets equipped with LoRa-radios and installing LoRa transceivers
in unmanned aerial vehicles (UAVs) will accelerate the localization
of the targets, probably unconscious. In this paper, the achievable
communication ranges of such links are theoretically and experimentally
evaluated by considering the possible positions of the helmet wearer
(standing or lying) on a flat field, representing a simple SaR scenario.
Simulations and experimental tests demonstrated that, for the standing
position, the ground-bounce multi-path produces strong fluctuations
of the received power versus the Tx-Rx distances. Such fluctuations
can be kept confined within 100 m from the target by lowering the
UAV altitude. Instead, for a more critical lying position, the received
power profile is monotonic and nearly insensitive to the posture.
For all the considered cases, the signal emitted by the body-worn
transceiver can be exploited to localize the helmet wearer based on
its strength, and it is theoretically detectable by the UAV radio
up to 5 km on flat terrain.

\medskip{}
\end{abstract}


\section{Introduction}\label{sec:Introduction}

\captionsetup[figure]{labelfont={default},labelformat={default},labelsep=period,name={Fig.}} \captionsetup[table]{labelfont={default},labelformat={default},labelsep=newline,name={TABLE},textfont={sc},justification=centering}The\let\thefootnote\relax\footnotetext{Work funded by the European Regional Development Fund under the Cooperation Programme Interreg V-A Italia Austria 2014-2020, ITAT3023, Smart Test for Alpine Rescue Technology START and by the European Regional Development Fund, Operational Program Investment for growth and jobs ERDF 2014-2020 under Project number ERDF1094, Data Platform and Sensing Technology for Environmental Sensing LAB-DPS4ESLAB.}\let\thefootnote\relax\footnotetext{Giulio Maria Bianco is with the Pervasive Electromagnetics Lab, University of Rome Tor Vergata, Rome, Italy, and also with the Center for Sensing Solution, EURAC research, Bolzano, Italy. Email: Giulio.Maria.Bianco@uniroma2.it (contact author)}\let\thefootnote\relax\footnotetext{Abraham Mejia-Aguilar is with the Center for Sensing Solutions, EURAC Research, Bolzano, Italy.}\let\thefootnote\relax\footnotetext{Gaetano Marrocco is with the Pervasive Electromagnetics Lab, University of Rome Tor Vergata, Rome, Italy.}
aim of search and rescue (SaR) operations is to localize a target
person and then provide assistance. Such procedures are very common
for both military and civilian purposes, particularly regarding the
identification of first responders during fires, earthquakes and floods
\cite{SaR_Fire_Flood,SaR_earthquakes} as well as in case of avalanches
and to locate lost hikers \cite{Accidentology}. Currently, ad-hoc
civilian SaR systems are used in mountain environments and rely on
radiofrequency devices: the avalanche beacons (also known as ARVAs
\cite{ARVA}) and the RECCO system \cite{RECCO}. However, the effectiveness
of those technologies is hampered by the low ranges, which spans from
about $60$ m (ARVA) to about $120$ m at most (RECCO). Moreover,
both devices cannot transmit any critical data about the user's health
status. Improvements could come from the internet of things (IoT)
systems exploiting the recently developed low-power wide-area networks
(LPWANs), such as LoRa, Sigfox and NB-IoT \cite{datarate}. Thanks
to their limited data rate, these networks can cover extended communication
distances while maintaining very low power consumption \cite{datarate}.
They can thus enable connectivity even in the harshest environments.

\begin{table}
\caption{Examples of relevant use cases of LoRa.\label{tab:Relevant-use-cases}}

\centering{}%
\begin{tabular}{|>{\raggedright}p{0.5cm}|>{\raggedright}p{3cm}||>{\raggedright}p{0.5cm}|>{\raggedright}p{3cm}|}
\hline 
\textbf{Ref.} & \textbf{Topic} & \textbf{Ref.} & \textbf{Topic}\tabularnewline
\hline 
\hline 
This work & Helmet-UAV off-body links for SaR & {[}13{]}, {[}14{]} & Terrestrial links for mountain SaR\tabularnewline
\hline 
{[}8{]} & Localization and tracking & {[}15{]} & Emergency communications\tabularnewline
\hline 
{[}9{]} & Remote health monitoring & {[}16{]} & Extend the SMS coverage to unconnected areas\tabularnewline
\hline 
{[}10{]} & Human activity recognition & {[}17{]} & Industrial monitoring\tabularnewline
\hline 
{[}11{]} & UAV communications for swarm configurations & {[}18{]} & Wireless underground sensor networks\tabularnewline
\hline 
{[}12{]} & Recovery of incapacitated UAVs & {[}19{]}, {[}20{]} & Smart cities\tabularnewline
\hline 
\end{tabular}%
\begin{comment}
Tabella di controllo per verificare che la tabella nbon commentata
sia corretta.
\begin{center}
\begin{tabular}{|>{\raggedright}p{0.5cm}|>{\raggedright}p{3cm}||>{\raggedright}p{0.5cm}|>{\raggedright}p{3cm}|}
\hline 
\textbf{Ref.} & \textbf{Topic} & \textbf{Ref.} & \textbf{Topic}\tabularnewline
\hline 
\hline 
This work & Helmet-UAV off-body links for SaR & \cite{Splitech,IoTJ} & Terrestrial links for mountain SaR\tabularnewline
\hline 
\cite{patienttracking} & Localization and tracking & \cite{Sisinni2020} & Emergency communications\tabularnewline
\hline 
\cite{LoRaHealth} & Remote health monitoring & \cite{Gimenez2020} & Extend the SMS coverage to unconnected areas\tabularnewline
\hline 
\cite{Shi2021} & Human activity recognition & \cite{Tran2020} & Industrial monitoring\tabularnewline
\hline 
\cite{Yuan18} & UAV communications for swarm configurations & \cite{Lin2020} & Wireless underground sensor networks\tabularnewline
\hline 
\cite{Catherwood2020} & Recovery of incapacitated UAVs & \cite{Aslam2020,Pham2020} & Smart cities\tabularnewline
\hline 
\end{tabular}
\par\end{center}
\end{comment}
\end{table}
 LoRa is one of the most investigated LPWAN technologies, as it was
proven to reach a $30$ km communication range by transmitting $25$
mW \cite{LoRarange}. Wearable LoRa devices are currently under study
for several applications ranging from tracking \cite{patienttracking}
to remote health monitoring \cite{LoRaHealth}. Relevant use cases
of LoRa are reported in Table \ref{tab:Relevant-use-cases}. LoRa
was also proposed for SaR applications where the communication range
and the power consumption are pivotal features. The authors recently
proposed a LoRa-based IoT system for mountain SaR operations \cite{IoTJ}.
The system\textbf{ }comprises\textbf{ }body-worn LoRa radios and a
range-based localization algorithm \cite{Lee2011} exploiting received
signal strength (RSS) measurements to localize the target when the
GPS signal is absent. Thanks to the low energy consumption of the
LoRa protocol, the radios can support the SaR operations for more
than $5$ hours and a half when they are fed with a typical battery
of $1100$ mAh capacity \cite{IoTJ}. Meanwhile, the target can be
localized even with a small number of measurements collected quickly
by employing classical range-based localization algorithms that exploit
the monotonic proportionality between the attenuation and the transmitter-receiver
distance, as \cite{Lee2011,Bianco21Multislope,Gholami13,Li2006}.
However, only terrestrial LoRa links have been considered so far,
whereas, in SaR scenarios, the terrestrial operations are complex
and slow. Therefore, LoRa-based SaR could greatly benefit from unmanned
aerial vehicles (UAVs) searching the target from the sky. Indeed,
by equipping a UAV with a LoRa receiver, many RSS measurements could
be easily collected in a short time over a wide area.

Most first responders and hikers wear a safety helmet during operations
and outdoor activities (like skiing or canyoneering), so a possible
placement of a LoRa wearable beacon is the helmet itself. Wearable
helmet-mounted antennas were proposed for disaster prevention \cite{disasterprevention}
and military forces \cite{radome}. The more common helmet antennas
are dipoles \cite{HelmetsSAR,HelmetSAR2}, patches \cite{radome},
and loops \cite{LoopHelmet}. Noticeably, as stated in Table \ref{tab:Relevant-use-cases},
a LoRa UAV-wearable device link \cite{Kachroo2019} has never been
neither characterized nor modeled. Moreover, all the helmet antennas
proposed so far are designed to work when the user is vigilant and
standing so that effects of the ground when the wearer is unconscious
or injured have not been addressed. Instead, if an accident happens,
the user is likely lying on the ground, eventually in the presence
of snow, and the helmet-UAV link is expected to be greatly affected
by the helmet wearer position.

This paper aims at evaluating the performance of radio-helmet-to-UAV
links when both standing and lying users are involved. Such off-body
links \cite{Ameloot21,Dumanli17} are analyzed for the purpose of
collecting through a UAV the LoRa RSS coming from a radio-helmet wearer
and activate a SaR procedure (the latter is outside of the scope of
this paper).

The modeling and experimentation are focused on line-of-sight (LoS)
scenarios in flat lands. However, despite the simple approximation,
the model can describe several SaR events, for example, \emph{i})
mountaineers in open environments hit by an avalanche, \emph{ii})
lost hikers in some natural parks after fires, \emph{iii}) injured
soldiers in desert areas, \emph{iv}) missing citizens and first responders
after cataclysms (as earthquakes, tornadoes, tsunami) that destroy
obstacles and create mostly flat and homogeneous zones. Within these
conditions, the maximum LoRa communication ranges and the RSS stability
versus the flight height and the UAV distance are here investigated.\textbf{
}The propagation model accounts\textbf{ }for several positions of
the helmet wearer (both standing and lying), the ground-bounce multi-path
and the LoRa transmission parameters (namely the variable sensitivity
of the receiver). Furthermore, the basic model here presented can
be extended to more complex environments by characterizing the radio
propagation of the considered site, either through measurement campaigns
\cite{IoTJ,Olasupo2019,Simunek13} or by ray-tracing simulations \cite{Cui2019,Safwat20}.

The paper is organized as follows. After the introduction of the most
relevant symbols (section \ref{sec:Symbols}), section \ref{sec:UAV-helmet-antenna-link}
details the problem formulation exploiting the link budget and simplified
modeling of the helmet transmitting antenna for some possible position
of a user with respect to the ground and the UAV receiver. A reliable
gain for the helmet antenna, suitable to any condition, is derived
for application to the link analysis in section \ref{sec:Link-evaluation}
through numerical simulations. An experimental campaign with a sensorized
UAV and a helmet embedding a LoRa radio is described in section \ref{sec:Measured-Signal-Strength},
and the measured data are compared, for corroboration, with the numerical
results. Finally, in section \ref{sec:Conclusions}, findings are
discussed, and conclusions are drawn.

\section{Symbols\label{sec:Symbols}}

The most relevant symbols used throughout the paper are here listed
for the reader's convenience in order of appearance.

\begin{tabular}{>{\raggedright}p{1cm}p{7cm}}
$G_{T}$ & radiation gain of the transmitting antenna\tabularnewline
\end{tabular}

\begin{tabular}{>{\raggedright}p{1cm}p{7cm}}
$G_{R}$ & radiation gain of the receiving antenna\tabularnewline
\end{tabular}

\begin{tabular}{>{\raggedright}p{1cm}p{7cm}}
$\varepsilon$ & electric permittivity: $\varepsilon_{0}\left[\varepsilon_{r}+j\cdot\textnormal{Im\ensuremath{\left(\unitfrac{\varepsilon}{\varepsilon_{0}}\right)}}\right]$\tabularnewline
\end{tabular}

\begin{tabular}{>{\raggedright}p{1cm}p{7cm}}
$\varepsilon_{0}$ & permittivity of vacuum\tabularnewline
\end{tabular}

\begin{tabular}{>{\raggedright}p{1cm}p{7cm}}
$\sigma$ & electrical conductivity\tabularnewline
\end{tabular}

\begin{tabular}{>{\raggedright}p{1cm}p{7cm}}
$P_{R}$ & power collected by the receiving antenna\tabularnewline
\end{tabular}

\begin{tabular}{>{\raggedright}p{1cm}p{7cm}}
$R$ & UAV-radio-helmet ground distance\tabularnewline
\end{tabular}

\begin{tabular}{>{\raggedright}p{1cm}p{7cm}}
$H$ & flying altitude of the UAV\tabularnewline
\end{tabular}

\begin{tabular}{>{\raggedright}p{1cm}p{7cm}}
$h$ & transmitter's height from ground\tabularnewline
\end{tabular}

\begin{tabular}{>{\raggedright}p{1cm}p{7cm}}
$t$ & terrain condition\tabularnewline
\end{tabular}

\begin{tabular}{>{\raggedright}p{1cm}p{7cm}}
$P_{T}$ & transmitting power\tabularnewline
\end{tabular}

\begin{tabular}{>{\raggedright}p{1cm}p{7cm}}
$\tau_{T}$ & power transfer coefficient of the transmitting antenna\tabularnewline
\end{tabular}

\begin{tabular}{>{\raggedright}p{1cm}p{7cm}}
$\chi$ & polarization loss factor\tabularnewline
\end{tabular}

\begin{tabular}{>{\raggedright}p{1cm}p{7cm}}
$PL$ & path loss\tabularnewline
\end{tabular}

\begin{tabular}{>{\raggedright}p{1cm}>{\raggedright}p{7cm}}
$F$ & path-gain factor\tabularnewline
\end{tabular}

\begin{tabular}{>{\raggedright}p{1cm}>{\raggedright}p{7cm}}
$r$ & ray-path\tabularnewline
\end{tabular}

\begin{tabular}{>{\raggedright}p{1cm}p{7cm}}
$d_{0}$ & reference distance of the log-distance path loss model\tabularnewline
\end{tabular}

\begin{tabular}{>{\raggedright}p{1cm}>{\raggedright}p{7cm}}
$n$ & path loss exponent\tabularnewline
\end{tabular}

\begin{tabular}{>{\raggedright}p{1cm}>{\raggedright}p{7cm}}
$\lambda$ & wavelength\tabularnewline
\end{tabular}

\begin{tabular}{>{\raggedright}p{1cm}p{7cm}}
$\Gamma_{T}$ & the module of the reflection coefficient of the transmitting antenna\tabularnewline
\end{tabular}

\begin{tabular}{>{\raggedright}p{1cm}p{7cm}}
$SF$ & spreading factor\tabularnewline
\end{tabular}

\begin{tabular}{>{\raggedright}p{1cm}p{7cm}}
$\rho e^{j\psi}$ & Fresnel reflection coefficient\tabularnewline
\end{tabular}

\begin{tabular}{>{\raggedright}p{1cm}p{7cm}}
$k_{0}$ & propagation constant in the vacuum: $2\pi\left(\lambda\right)^{-1}$\tabularnewline
\end{tabular}

\begin{tabular}{>{\raggedright}p{1cm}p{7cm}}
$\varphi$ & incidence angle\tabularnewline
\end{tabular}

\begin{tabular}{>{\raggedright}p{1cm}p{7cm}}
$\hat{\rho}_{T}$ & polarization versor of the transmitting antenna\tabularnewline
\end{tabular}

\begin{tabular}{>{\raggedright}p{1cm}p{7cm}}
$\hat{\rho}_{R}$ & polarization versor of the receiving antenna\tabularnewline
\end{tabular}

\section{Helmet-to-UAV link \label{sec:UAV-helmet-antenna-link}}

\begin{figure}
\begin{centering}
\includegraphics[width=8cm]{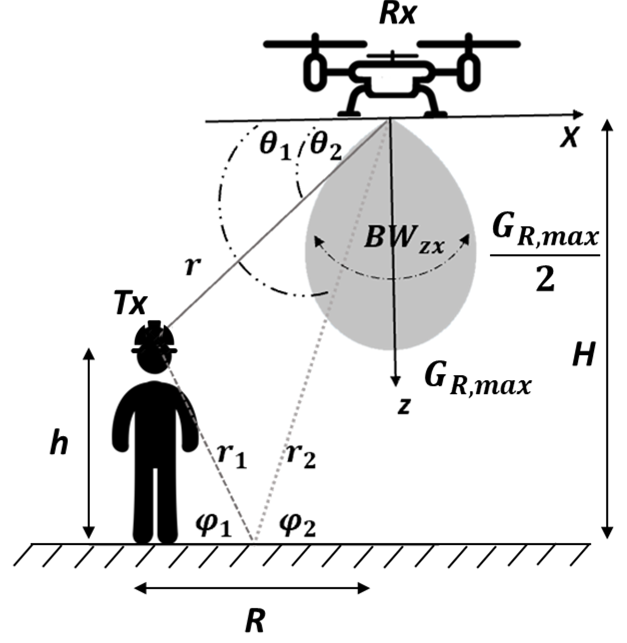}
\par\end{centering}
\caption{Sketch of the radio-helmet-to-UAV link and relevant parameters.\label{fig:Two_ray_gain}}
\end{figure}

\begin{figure}
\begin{centering}
\includegraphics[width=2.5cm]{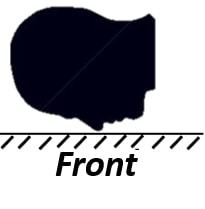}(a)\includegraphics[width=2.5cm]{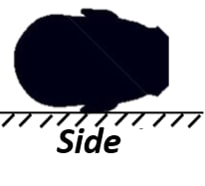}(b)\includegraphics[width=2.5cm]{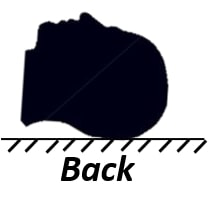}(c)
\par\end{centering}
\caption{The three possible positions of a lying user (a) on the front, (b)
on the side, and (c) on the back. \label{fig:boundaries}}
\end{figure}

With reference to Fig. \ref{fig:Two_ray_gain}, the helmet-to-UAV
link involves a LoRa transmitter (working in the $863-873$ MHz LoRa
band \cite{LoRaBand}) connected to a helmet antenna having radiation
gain $G_{T}$ and a UAV equipped with a receiving antenna of gain
$G_{R}$. The user is assumed in four possible positions: standing,
lying on the front, side-lying, lying on the back (Fig. \ref{fig:boundaries}).
Two reference boundary ground conditions are considered: a perfect
electric conductor (PEC) ground (approximating a land covered by snow)
and extremely dry terrain {[}i.e., rocks; permittivity $\varepsilon=\varepsilon_{0}\left(4.8-0.4j\right)$,
electrical conductivity $\sigma=10^{-4}$ S/m \cite{dry terrain}{]}.
The behavior of any real terrain is expected to be comprised between
the aforementioned two extreme boundary conditions.

The helmet-to-UAV link is hence described by the power $P_{R}$ collected
by the receiver on the UAV according to the following link budget
(all terms in dB scale) 

\begin{equation}
\begin{array}{c}
P_{R}\left(R,H,h,t\right)=P_{T}+G_{T}\left(R,H,h,t\right)+G_{R}\left(R,H,h\right)\\
+\tau_{T}(h,t)+\chi\left(R,H,h,t\right)-PL\left(R,H,h\right)+F(R,H,h,t)
\end{array},\label{eq:linkbudget}
\end{equation}
where $P_{T}$ is the transmitting power, $\chi$ is the polarization
loss factor, $h$ is the transmitter height from the ground, \emph{H}
is the flying altitude of the UAV, \emph{R }is the UAV's ground distance
from the target, and $t$ is the terrain condition (dry or wet). The
log-distance \emph{PL} (path loss) expression is $PL\left(r\right)=PL\left(d_{0}\right)+10\cdot n\cdot\log_{10}\left(\nicefrac{r}{d_{0}}\right)$,
being $d_{0}$\emph{ }a reference distance, \emph{n }the path loss
exponent and $r$ the ray-path. Since LoS ground-to-air channels of
low-altitude UAVs can be approximated by the free-space link \cite{UAV_GtoA_Propagation},
the conditions $d_{0}=\unitfrac{\lambda}{4\pi}$ and $n=2$ are assumed.
Being $\Gamma_{T}$ the module of the reflection coefficient of the
transmitting antenna, $\tau_{T}=(1-\Gamma_{T}^{2})$ is the corresponding
power transfer coefficient. The receiving antenna is instead considered
perfectly matched to the receiver. The theoretical maximum communication
distance is evaluated by imposing $P_{R}$ equal to the receiver's
sensitivity, which depends on the bandwidth and the spreading factor
(\emph{SF}) of the LoRa signal as in Table \ref{tab:Sensitivitiy-and-bit}.
Based on \cite{IoTJ} and \cite{TransmissionParameersLoRa}, a signal
bandwidth of $125$ kHz and a variable $SF$ value are considered
throughout the paper. 
\begin{table}
\caption{Sensitivity of the LoRa SX1276 transceiver for different \emph{SF}
values at bandwidth $125$ kHz. \cite{LoRaSensitivities}\label{tab:Sensitivitiy-and-bit}}

\centering{}%
\begin{tabular}{|c|c|}
\hline 
\textbf{Spreading Factor} & \textbf{Sensitivity {[}dBm{]}}\tabularnewline
\hline 
\hline 
12 & -136\tabularnewline
\hline 
11 & -133\tabularnewline
\hline 
10 & -132\tabularnewline
\hline 
9 & -129\tabularnewline
\hline 
8 & -126\tabularnewline
\hline 
7 & -123\tabularnewline
\hline 
\end{tabular}
\end{table}
Finally, the term \emph{F} is the path-gain factor that accounts for
the multi-path in the case of a standing user through the flat-earth
two-ray propagation model \cite{UAV_GtoA_Propagation,CoverageMaps}.
With reference to the geometrical parameters in Fig. \ref{fig:Two_ray_gain}

{\small{}
\begin{equation}
F=20\log_{10}\left[\left|1+\rho e^{j\psi}\sqrt{\frac{G_{T}\left(\theta_{1}^{'}\right)G_{R}\left(\theta_{1}\right)}{G_{T}\left(\theta_{2}^{'}\right)G_{R}\left(\theta_{2}\right)}}e^{-jk_{0}\left(r-r_{1}-r_{2}\right)}\right|\right],\label{eq:path_gain}
\end{equation}
}where $k_{0}$ is the free-space propagation constant, $\rho e^{j\psi}$
is the Fresnel reflection coefficient, $\theta_{2}$ identifies the
direction of the direct path, and the antennas' gains are in linear
scale. By assuming a smooth ground, then $\varphi_{1}=\varphi{}_{2}=\varphi$.
The Fresnel reflection coefficient for an electromagnetic wave polarized
parallelly to the ground is evaluated from the permittivity of the
land itself as \cite{CoverageMaps}

\begin{equation}
\rho e^{j\psi}=\frac{\sin\varphi-\sqrt{\nicefrac{\varepsilon}{\varepsilon_{0}}-\cos^{2}\varphi}}{\sin\varphi+\sqrt{\nicefrac{\varepsilon}{\varepsilon_{0}}-\cos^{2}\varphi}}.
\end{equation}

For PEC reflecting surfaces, then $\rho e^{j\psi}=-1$. If the user
is lying, the antenna is on the ground, the multi-path disappears,
\emph{F} is dropped out from (\ref{eq:linkbudget}), and the ground
effect is accounted for in the gain of the helmet antenna as described
in the next section.

The link budget in (\ref{eq:linkbudget}) is here parametrized regarding
the user's position (through the transmitter gain) and the features
of the receiving antenna.

\subsection{Model of the Rx antenna}

The UAV is hereafter assumed to be equipped with a circularly polarized
(CP) patch as in \cite{UAVFootprint}. The gain $G_{R}$ (in linear
scale) can be roughly approximated with an ellipsoid having rotational
symmetry \cite{ellipsoid} whose elliptic section can be expressed
in polar coordinates (w.r.t. Fig. \ref{fig:Two_ray_gain})

\begin{equation}
G_{R}\left(\theta\right)=\frac{2a_{z}^{2}a_{\xi}\sin\theta}{a_{z}^{2}\left(\cos\theta\right)^{2}+a_{\xi}^{2}\left(\sin\theta\right)^{2}}.\label{eq:ellipse}
\end{equation}

The patch is assumed to be the pole and lie on the $\left(x,y\right)$
plane. The ellipse axes hence are

\begin{equation}
a_{z}=\frac{G_{R,max}}{2}
\end{equation}

\begin{equation}
a_{\xi}=a_{z}\cot\left(\frac{BW_{z\xi}}{2}\right),
\end{equation}
where \{$G_{R,max}$, $BW_{z\xi}$\} are the maximum gain of the antenna
and its half-power beamwidth, respectively.

\subsection{Model of the Tx antenna}

\begin{figure}
\begin{centering}
\includegraphics[width=4.5cm]{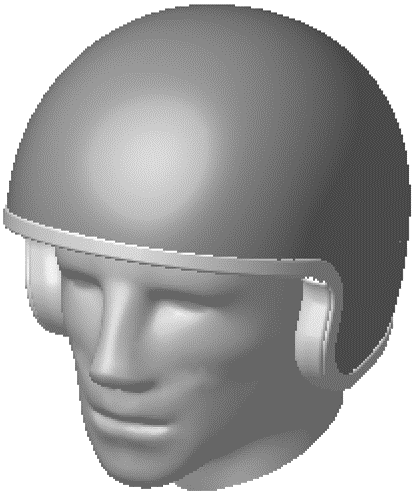}
\par\end{centering}
\begin{centering}
(a)
\par\end{centering}
\begin{centering}
\includegraphics[width=3cm]{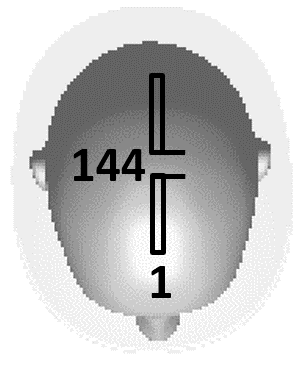} $\qquad$\includegraphics[width=3cm]{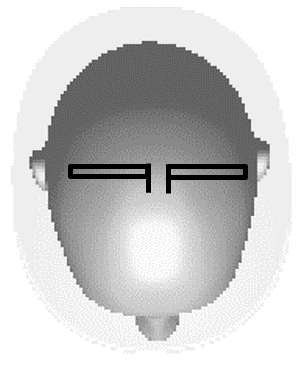}
\par\end{centering}
\begin{centering}
(b)$\qquad\qquad\qquad\qquad\qquad($c)
\par\end{centering}
\caption{(a) Helmet and head numerical phantoms. The flat dipole (size in mm)
along (b) the\textbf{ }medial sagittal and (c) the coronal planes.
\label{fig:Helmet-and-head}}
\end{figure}

The choice of the reference helmet antenna could, in principle, consider
both a linearly-polarized (LP) or a CP device as for the receiver
onboard the UAV. A CP antenna would permit to minimize the interference
due to the ground since the reflection would invert the polarization
verse so that the reflected field is filtered out by the receiver
antenna having opposite polarization. However, using a CP antenna
on the helmet too would introduce some drawbacks since the polarization
of the helmet antenna could be seen as reversed by the receiver because
of the unpredictable Tx-Rx mutual positions during the UAV flight
and the actual posture of the user. Sharp polarization mismatches
could occur hence preventing the establishment of the communication.
For the sake of generality, the transmitting antenna is assumed to
be a flat dipole (copper trace $35$ \textgreek{m}m thick) wrapped
onto the top of the helmet (Fig. \ref{fig:Helmet-and-head}) according
to two possible arrangements (w.r.t. the anatomical planes): along
the median sagittal plane (hereafter \emph{sagittal dipole}), and
the coronal plane (\emph{coronal dipole}). As shown next in section
\ref{subsec:Polarization-Loss-Factor}, this arrangement guarantees
a reliable polarization matching in most conditions.

\begin{figure}
\begin{centering}
\includegraphics[width=4cm]{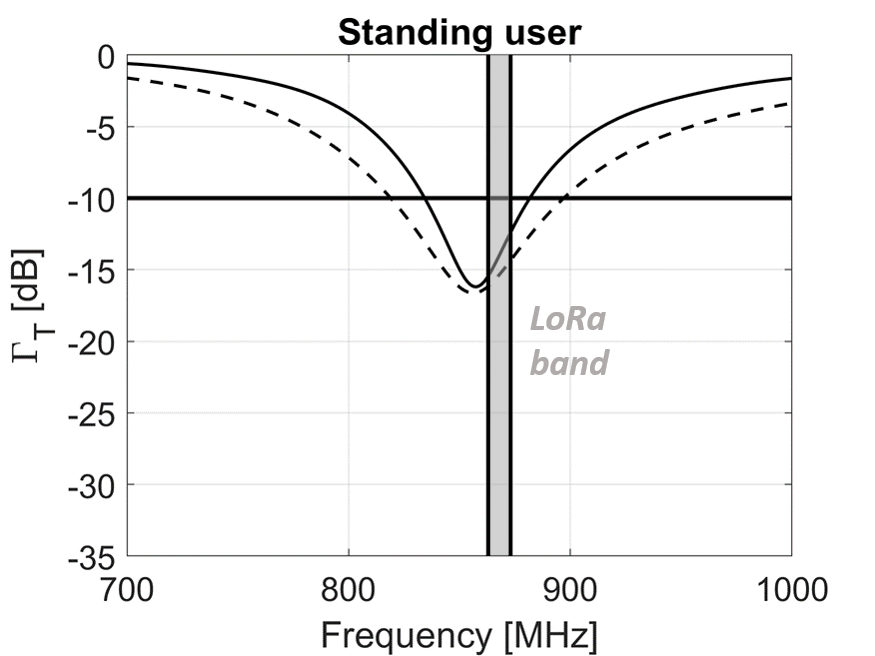}\includegraphics[width=4cm]{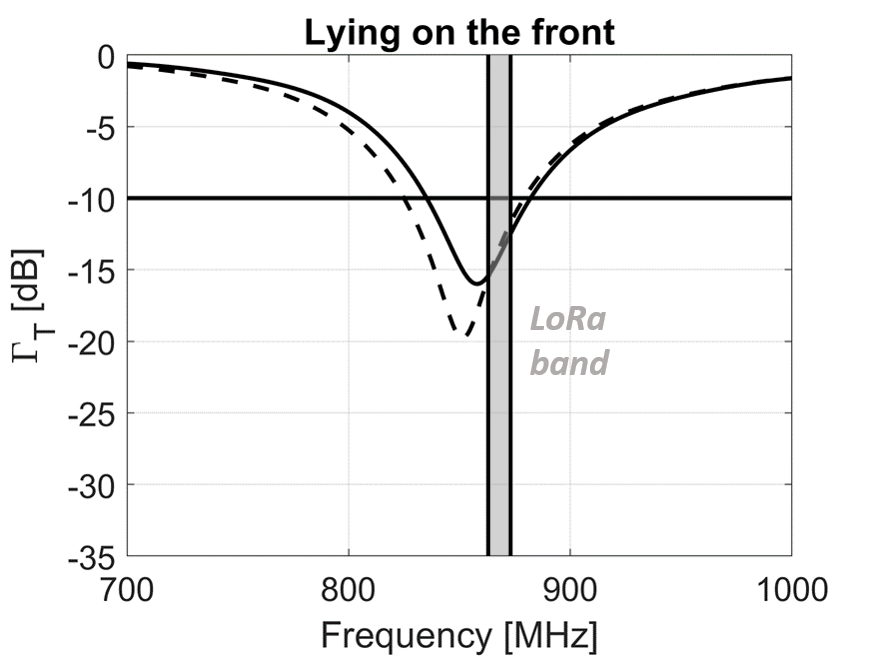}
\par\end{centering}
\begin{centering}
\includegraphics[width=4cm]{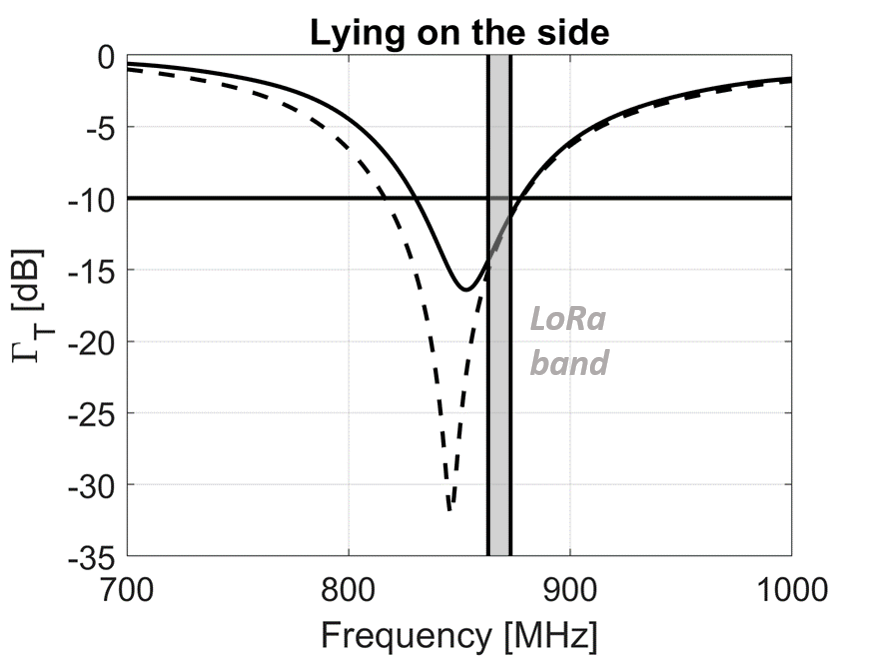}\includegraphics[width=4cm]{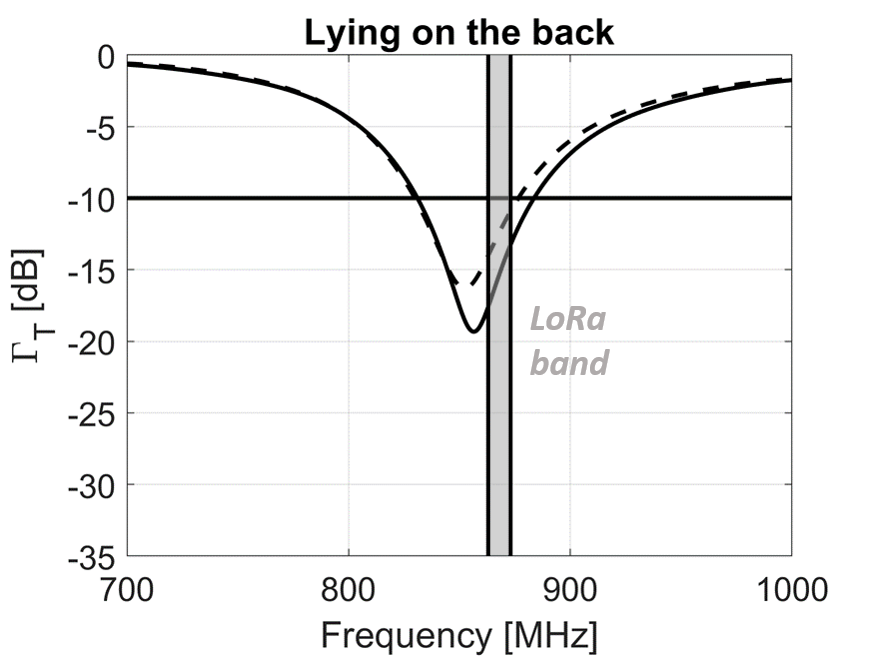}
\par\end{centering}
\smallskip{}

\begin{centering}
\includegraphics[width=5cm]{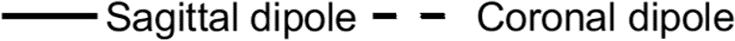}
\par\end{centering}
\caption{Simulated reflection coefficients of the helmet antennas in Fig. \ref{fig:Helmet-and-head}
in four user positions over PEC terrain. The LoRa 863-873 MHz band
is highlighted in grey. \label{fig:Reflecion-coefficient-S11}}
\end{figure}

The gain $G_{T}$ of the transmitting antenna is dependent on the
working conditions. In particular, when the user is lying, the interaction
with the ground is expected to disturb the radiation pattern strongly.
To simplify the evaluation of the link budget in (\ref{eq:linkbudget}),
the transmitting antenna is taken into account through an equivalent
uniform gain pattern whose value is derived from statistical analyses
involving the user's possible positions. For this purpose, the radiation
performances of all the combinations of antennas and positions are
numerically evaluated utilizing the software CST Microwave Studio
Suite 2018. The numerical model\footnote{Model available at https://grabcad.com/library/helmet-184.}
employed for the simulations includes:
\begin{itemize}
\item a homogeneous numerical phantom of a human head with average relative
permittivity and conductivity $\varepsilon_{r}=42.7,\ \sigma=0.99$
S/m \cite{headSAR} respectively;
\item a lossless foamed plastic ($\varepsilon_{r}=1.5$ \cite{foamcostant})
shell with the dimensions of a typical helmet for mountaineering (derived
from the Vayu 2.0 model by Salewa, having perimeter length of $63$
cm);
\item ground modeled as a conducting plate.
\end{itemize}
The dipoles were preliminarily tuned at $868$ MHz in the standing
user case so that the reflection coefficient is $\varGamma_{T}\leq-10$
dB in the whole LoRa band. The proximity with the PEC ground in the
case of a lying user produces a moderate impedance mismatch (Fig.
\ref{fig:Reflecion-coefficient-S11}), nevertheless preserving the
same maximum value as before. If the terrain is dry, the reflection
coefficients are less affected by the helmet position and are more
similar to the standing user case.

\begin{figure}
\begin{centering}
\textbf{\large{}Standing user}{\large\par}
\par\end{centering}
\begin{centering}
\includegraphics[width=4cm]{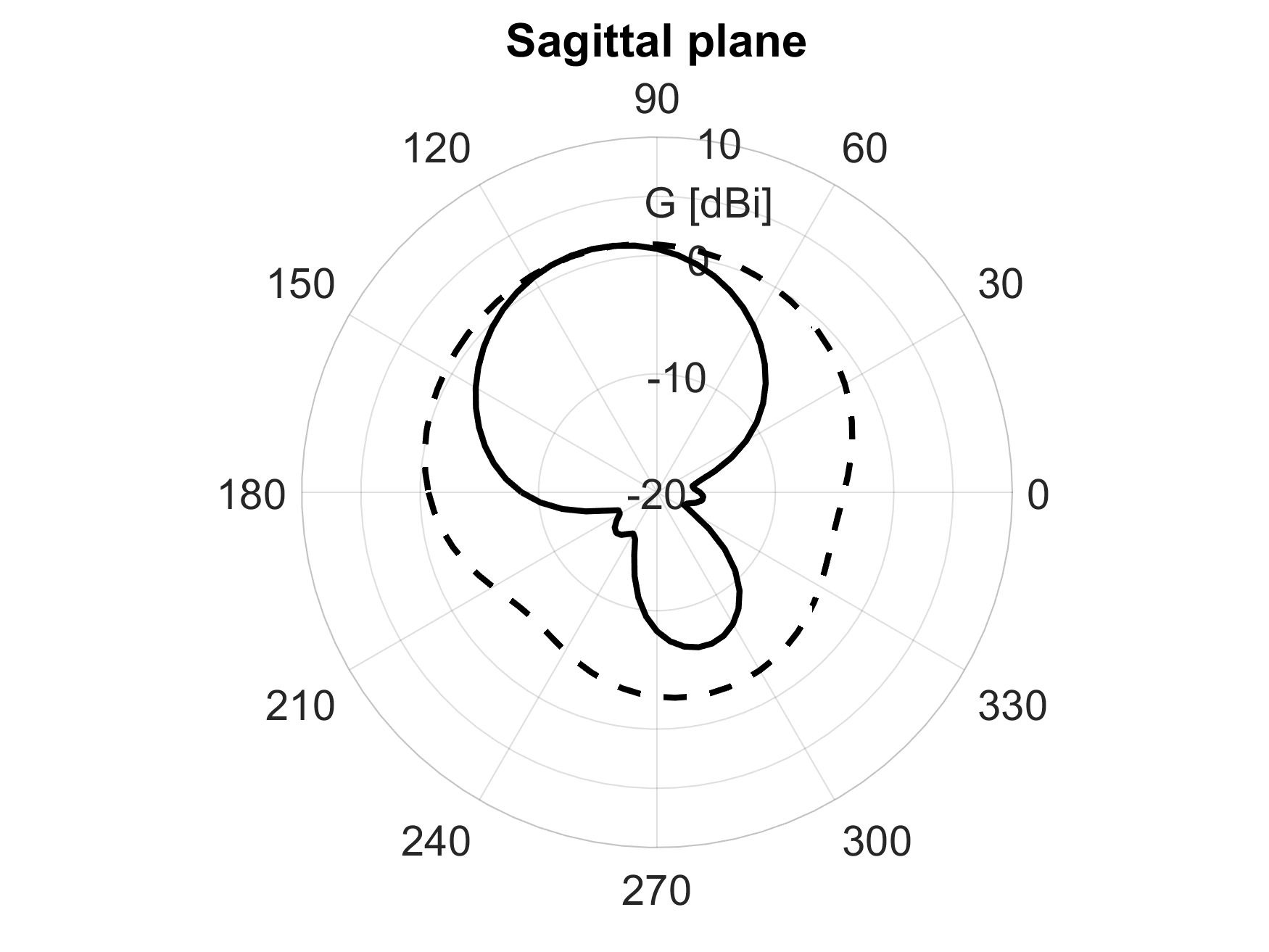}\includegraphics[width=4cm]{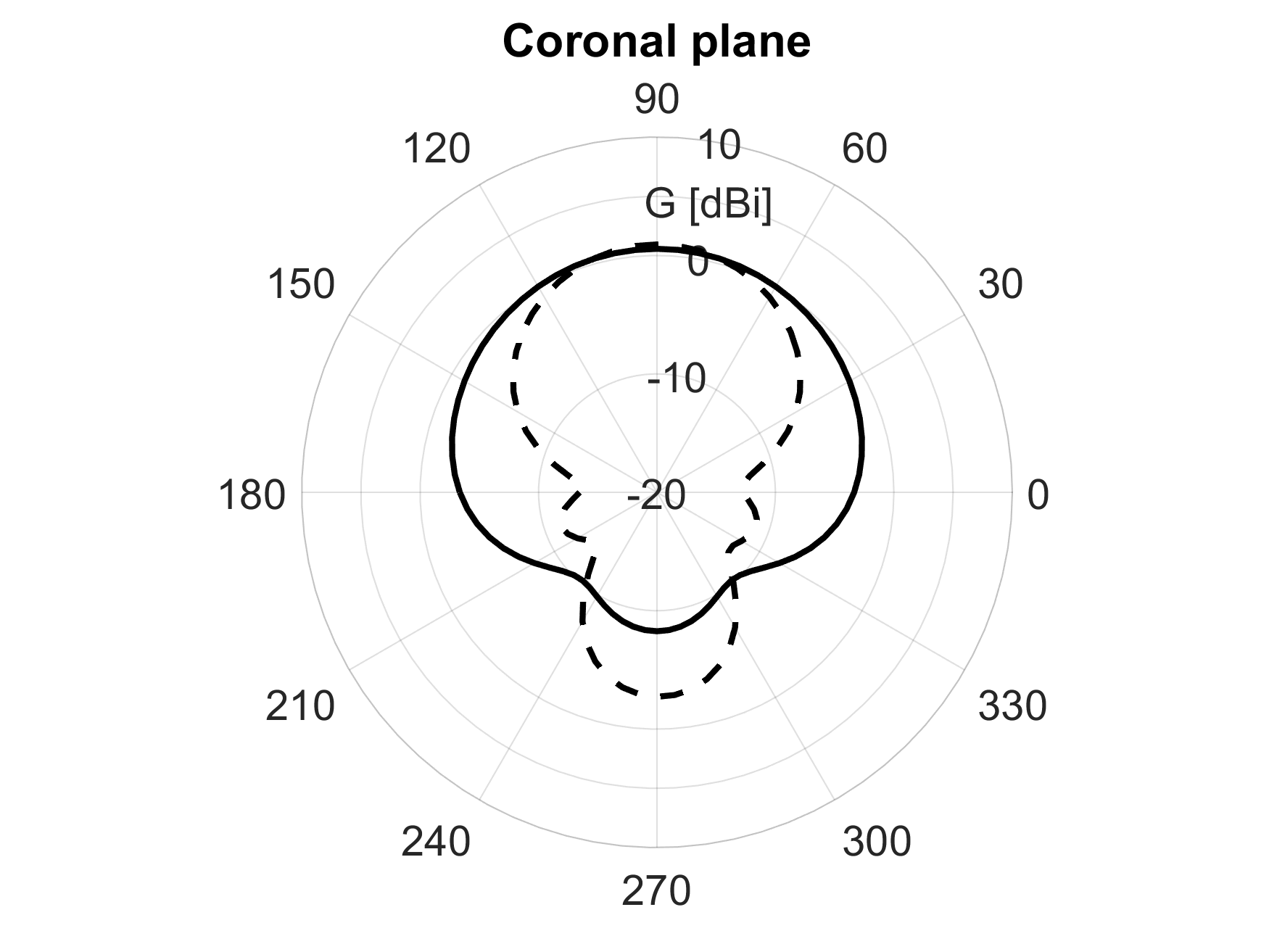}
\par\end{centering}
\begin{centering}
\textbf{\large{}Lying user}{\large\par}
\par\end{centering}
\begin{centering}
{\large{}Front}{\large\par}
\par\end{centering}
\begin{centering}
\includegraphics[width=4cm]{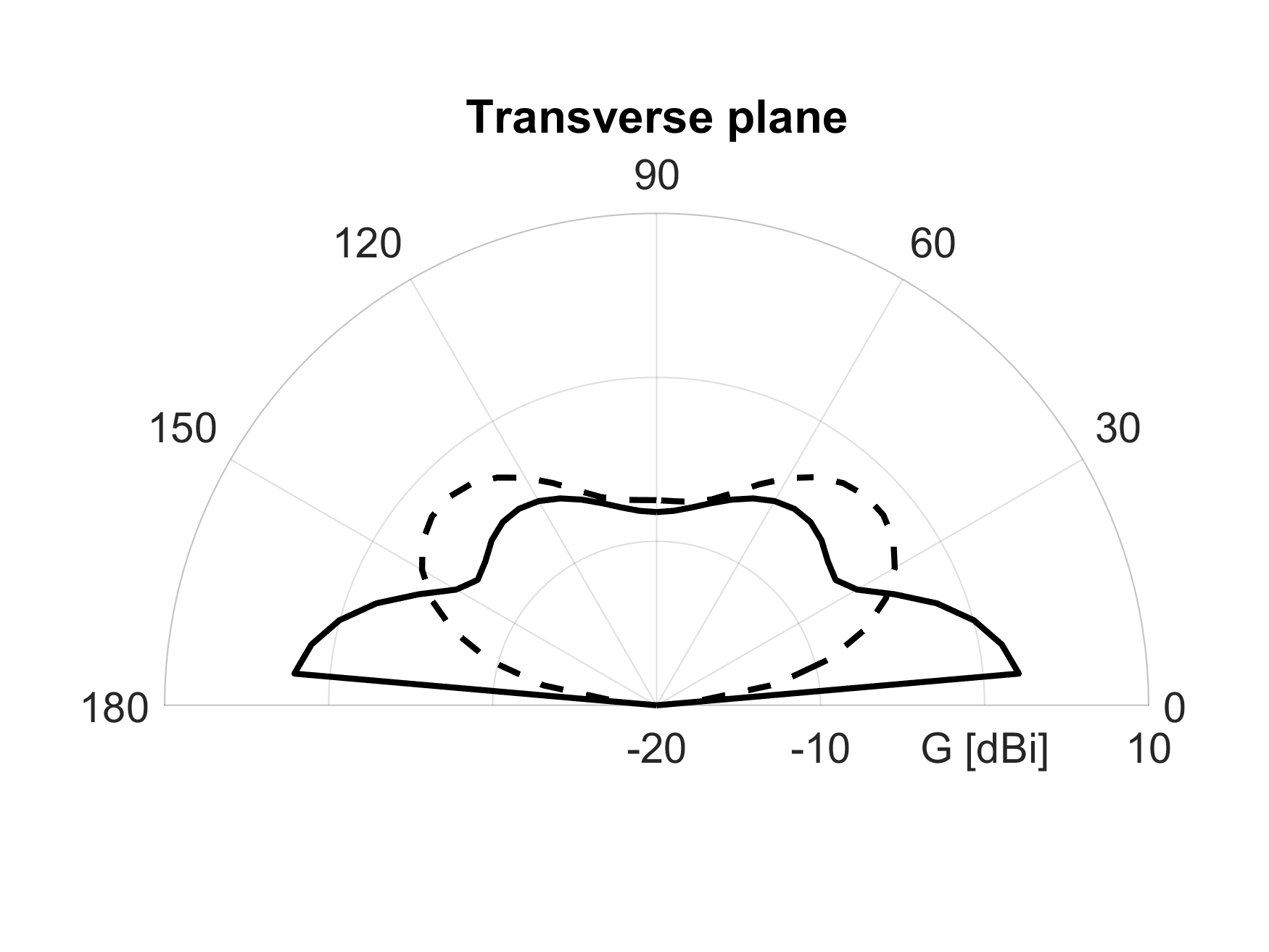}\includegraphics[width=4cm]{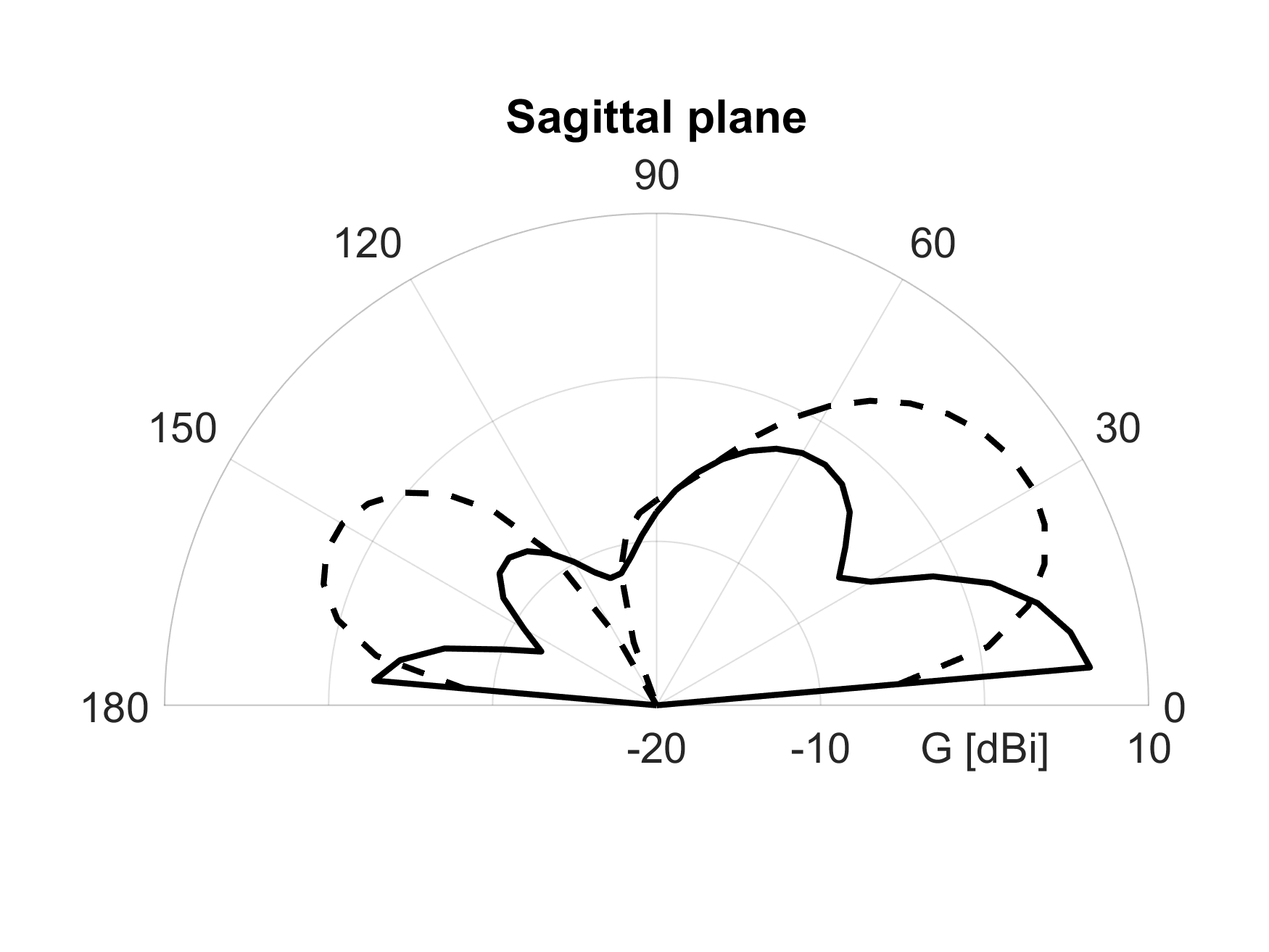}
\par\end{centering}
\begin{centering}
{\large{}Side}{\large\par}
\par\end{centering}
\begin{centering}
\includegraphics[width=4cm]{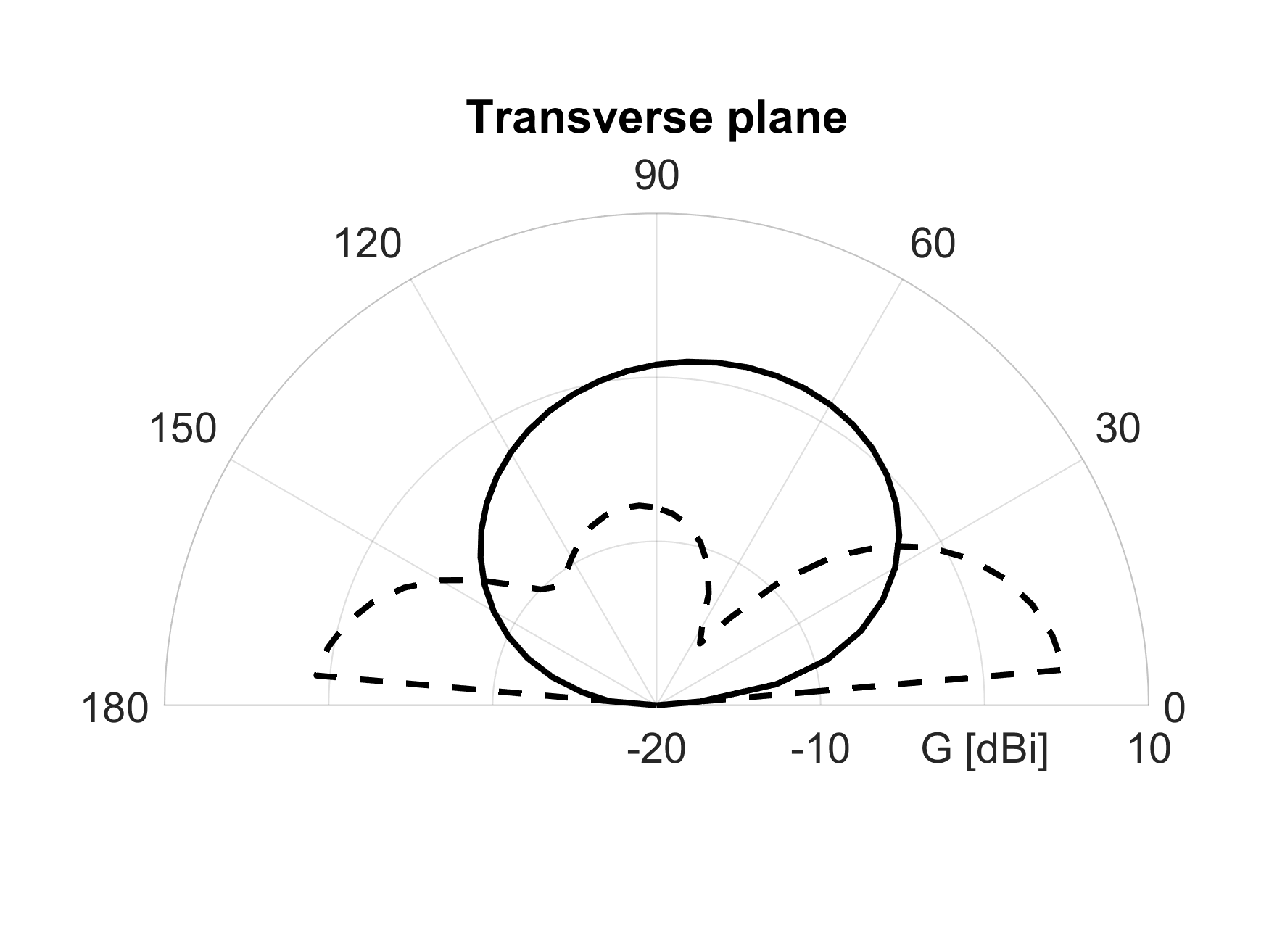}\includegraphics[width=4cm]{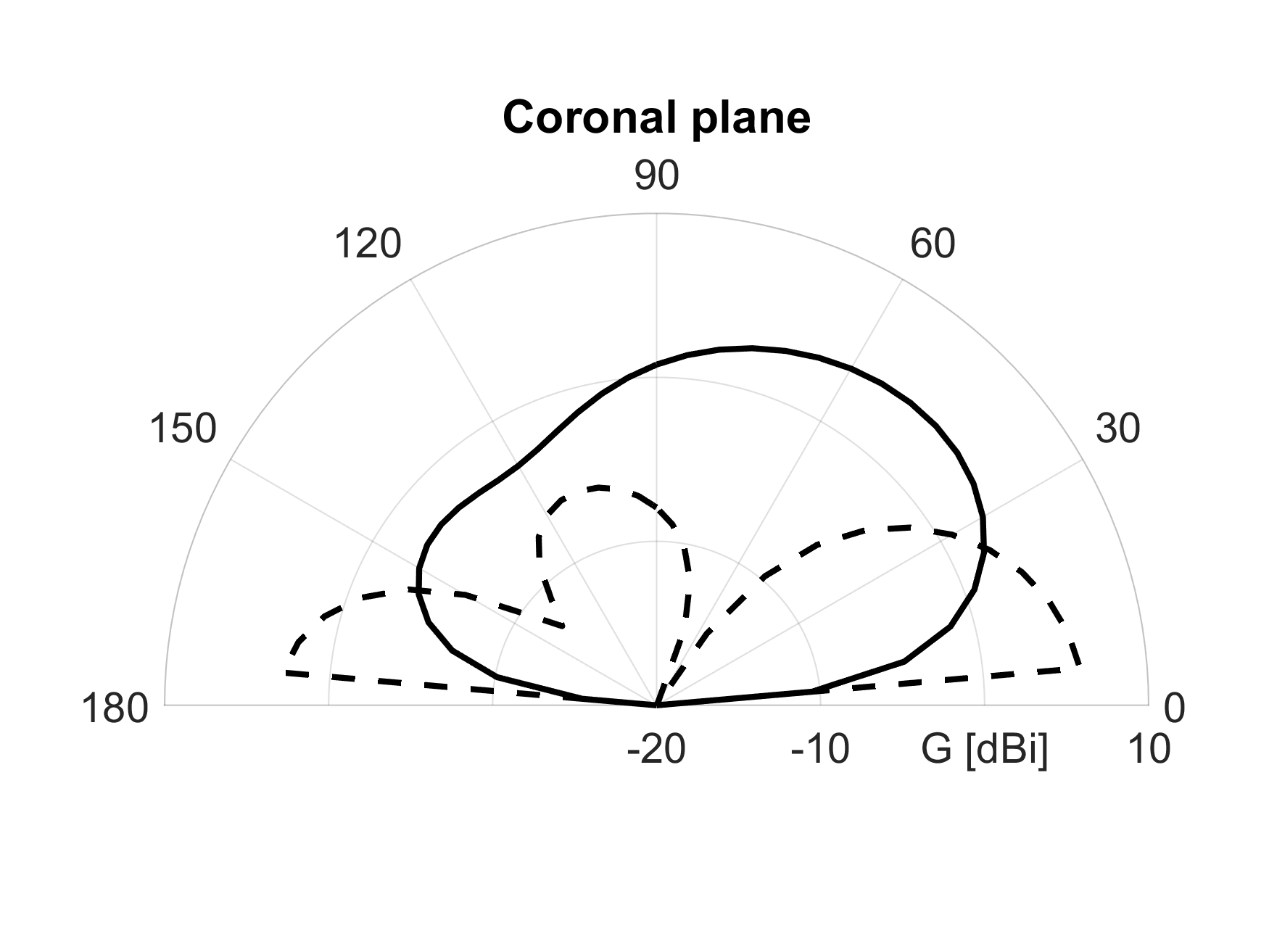}
\par\end{centering}
\begin{centering}
{\large{}Back}{\large\par}
\par\end{centering}
\begin{centering}
\includegraphics[width=4cm]{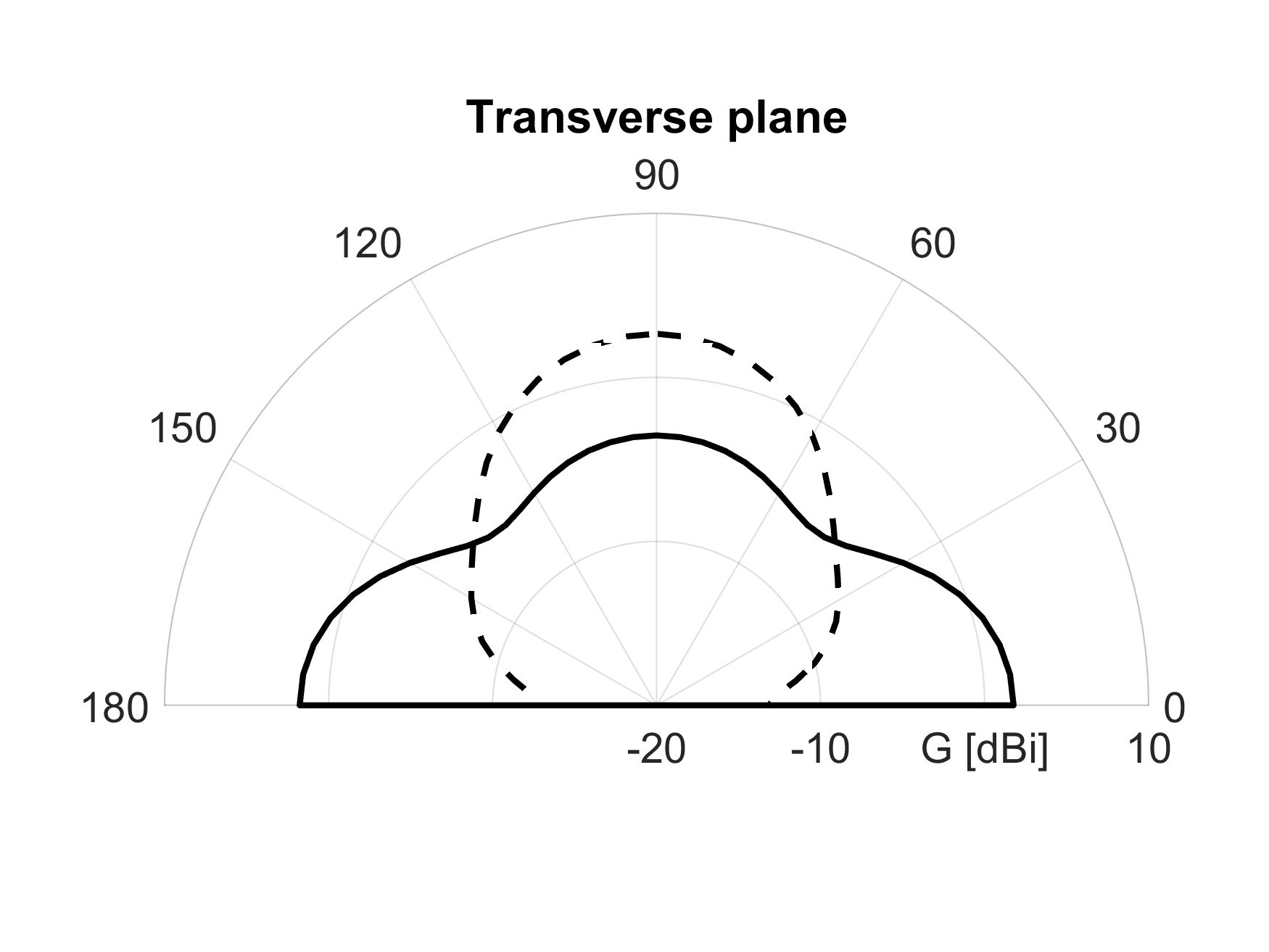}\includegraphics[width=4cm]{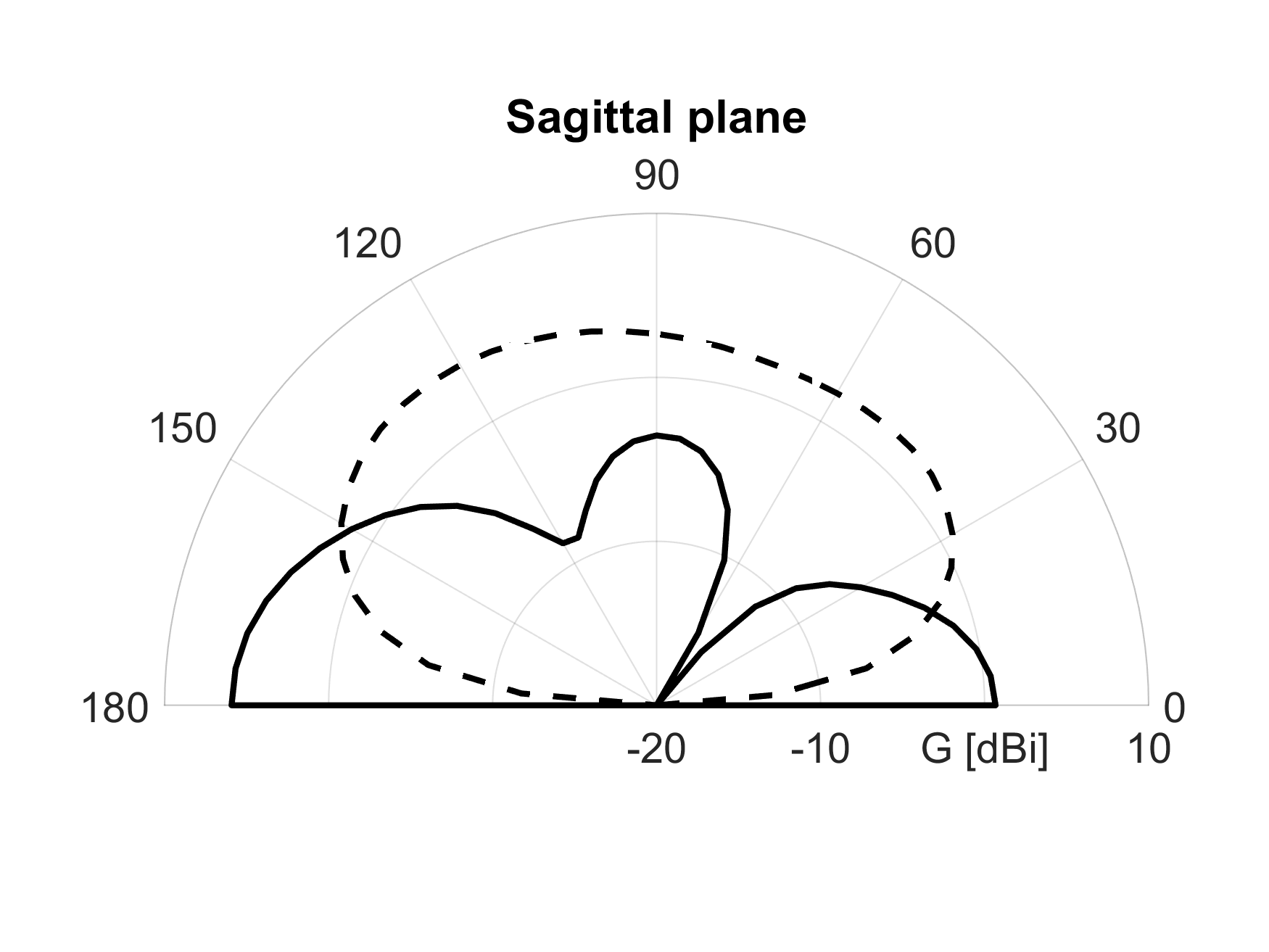}
\par\end{centering}
\smallskip{}

\begin{centering}
\includegraphics[width=5cm]{Images/legend}
\par\end{centering}
\caption{Simulated radiation gain of the helmet antennas (sagittal and coronal
dipoles) over planes passing through the zenith.\textbf{ }In the case
of a lying user, a PEC ground is assumed. The dry terrain case returns
the same patterns with an offset of $-2$ dB.\textbf{ }\label{fig:Gains}}
\end{figure}

 By referring to the radiation gain patterns at the zenith (Fig. \ref{fig:Gains}),
it is clear that the two antenna layouts behave similarly for a standing
user. In contrast, relevant differences are visible in the case of
a lying user, with the maximum radiation occurring either along the
zenith or the horizon depending on the position of the helmet.

\begin{figure}
\begin{centering}
\begin{tabular}{cc}
\includegraphics[width=4cm]{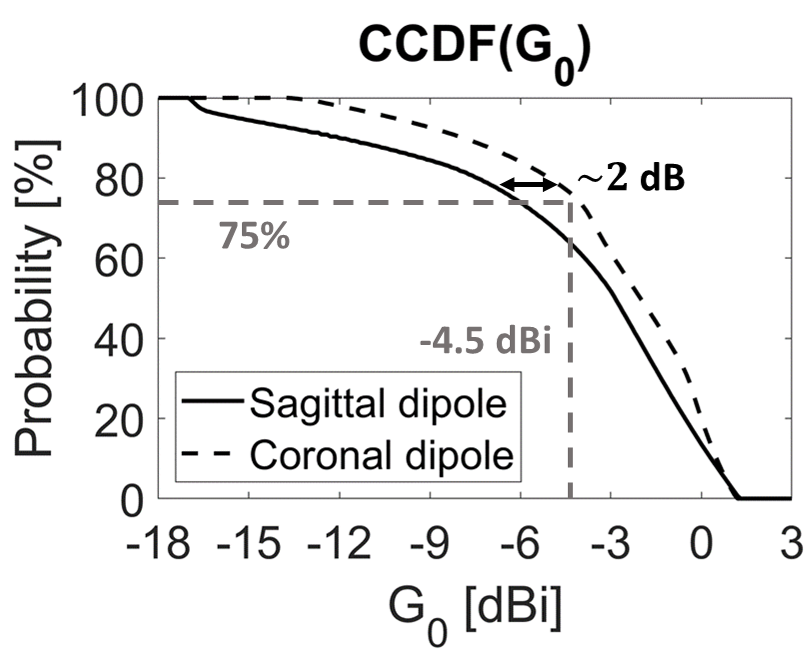} & \includegraphics[width=4cm]{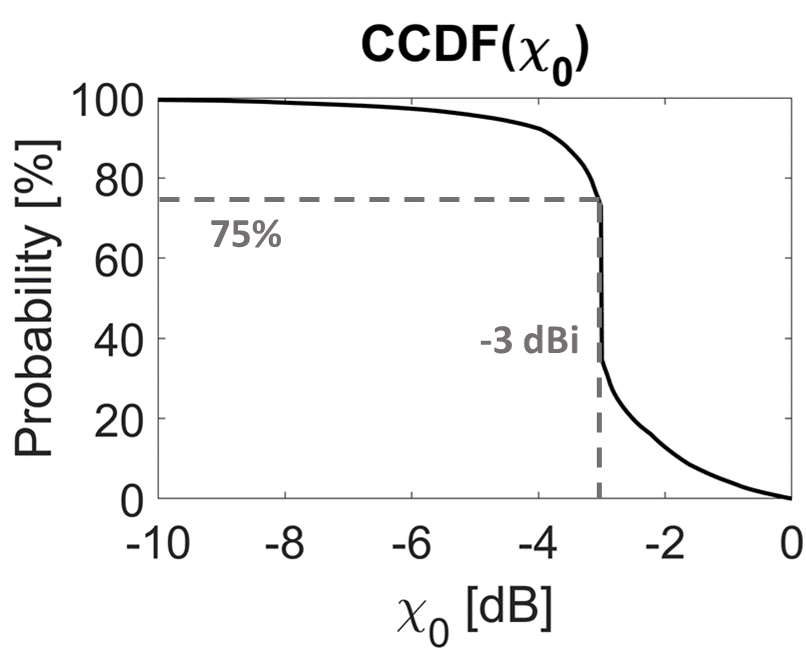}\tabularnewline
(a) & (b)\tabularnewline
\end{tabular}
\par\end{centering}
\caption{Complementary cumulative distribution functions (a) $CCDF(G_{0})$
for the sagittal and coronal\textcolor{black}{{} dipoles by taking into
account the data }of all the three lying positions of the user on
a PEC ground (Fig. \ref{fig:boundaries}), and (b) $CCDF(\chi_{0})$
for the coronal dipole when considering both the standing and the
lying postures. \label{fig:Complementary-Cumulative-Distrib}}
\end{figure}

 The most appropriate antenna arrangement is the one maximizing the
uniformity of the radiation pattern in the upper half-space in all
the considered user configurations, especially in the lying ones.
The complementary cumulative distribution function (CCDF) for $G_{T}>G_{0}$
in the whole half-space\footnote{$CCDF(G_{0})$ is the percentage of the half-space where the gain
of the antenna in all the three lying configurations is more than
$G_{0}$.} is used as a metric {[}Fig. \ref{fig:Complementary-Cumulative-Distrib}(a){]}.
Accordingly, the coronal dipole outperforms the sagittal one. By considering,
for instance, the threshold value of probability $75\%$, the corresponding
lower bound gain $G_{0}\left(CCDF=75\%\right)$ of the coronal dipole
is about $2$ dB higher than the corresponding value of the sagittal
dipole.

In conclusion, the coronal dipole is hereafter considered the transmitting
antenna whose equivalent gain is $G_{T}=G_{0}\left(CCDF=75\%\right)$,
keeping the different values for standing and lying cases and kind
of terrain, as summarized in Table \ref{tab:The-lower-bound-gains}.
\begin{table}
\caption{Selected equivalent Tx gain $G_{0}\left(CCDF=75\%\right)$ for different
user and terrain conditions.\label{tab:The-lower-bound-gains}}

\centering{}%
\begin{tabular}{|c|c|c|}
\hline 
\textbf{Standing user} & \multicolumn{2}{c|}{$-4.1$ dBi}\tabularnewline
\hline 
\multirow{2}{*}{\textbf{Lying user}} & Wet terrain & Dry terrain\tabularnewline
\cline{2-3} \cline{3-3} 
 & $-4.5$ dBi & $-6.5$ dBi\tabularnewline
\hline 
\end{tabular}
\end{table}

\subsection{Polarization Loss Factor \label{subsec:Polarization-Loss-Factor}}

Like the antennas' gains, the polarization loss factor heavily varies
with the helmet wearer's position and the terrain condition. The loss
factor is evaluated from the polarization versors of the transmitter
$\left(\hat{\rho}_{T}\right)$ and the receiver $\left(\hat{\rho}_{R}\right)$
as

\begin{equation}
\chi=\left|\hat{\rho}_{T}\bullet\hat{\rho}_{R}^{*}\right|^{2}\label{eq:PLF}
\end{equation}
$\left|\cdot\right|$, $^{*}$ and $\bullet$ being the module, the
complex conjugate, and the inner product operators, respectively.

To model the polarization effects, the polarization of the receiving
CP patch is assumed constant and equal to $\hat{\rho}_{R}=\sqrt{2^{-1}}\left(1+j\right)$,
whereas the versor of the transmitting antenna is numerically evaluated
when considering all the position-terrain combinations and all the
possible observation angles {[}Fig. \ref{fig:Complementary-Cumulative-Distrib}(b){]}.
From the simulations, $\chi\geq-3$ dB in $75\%$ of the cases and,
accordingly, the assumption $\chi=-3$ dB is considered for the next
evaluation of the received power $P_{R}$.

\section{Link evaluation\label{sec:Link-evaluation}}

Following the above framework, the radio-helmet-to-UAV link is numerically
evaluated to predict the maximum achievable communication range and
analyze the two-ray interference through coverage maps. The communication
range is expressed as the radio-helmet-UAV ground distance since it
is a crucial parameter for most SaR operations. The user-UAV link
must be as monotonic as possible to apply range-based location algorithms
exploiting the RSS effectively.

The RSS is evaluated according to (\ref{eq:linkbudget}). The parameters
of the receiving antenna to derive $G_{R}(\theta)$ are referred to
the Keonn Advantenna-p11 ($BW_{z\xi}=100^{\textnormal{o}}$, $G_{R,max}=3.2$
dBi), which is then used for the experimentation described in the
next section. The received power at the UAV side is evaluated for
a variable flight altitude $5$ m $\leq H\leq120$ m of the UAV.\footnote{\emph{H} = 120 m is the highest altitude of a UAV that is compliant
with EU regulations on civilian applications (open category). \cite{UAV_open_category}} The corresponding maximum radio range $R_{max}$, wherein the received
power $P_{R}$ equals the sensitivity of the receiver, is derived.
The transmitter must moreover comply with the ERP (effective radiated
power) regulation for LoRa systems. By enforcing the EU constraint
\cite{EIRP_limit_LoRa} over the maximum LoRa irradiation (i.e., ERP
$=14$ dBm at $868$ MHz), the maximum transmitter power to insert
in (\ref{eq:linkbudget}) is $P_{T}=14$ dBm. All computations refer
to an $h=1.7$ m tall user. Finally, from Fig. \ref{fig:Reflecion-coefficient-S11},
$\tau_{T}(h,t)\simeq\tau_{T}=-0.01$ dB. The values of the parameters
employed in the numerical simulations are resumed in Table \ref{tab:The-lower-bound-gains}
and Table \ref{tab:Parameters-considered-in}. 
\begin{table}
\caption{Considered values for the numerical RSS evaluation.\label{tab:Parameters-considered-in}}

\centering{}%
\begin{tabular}{|l|l||l|l|}
\hline 
\textbf{Parameter} & \textbf{Value} & \textbf{Parameter} & \textbf{Value}\tabularnewline
\hline 
\hline 
$P_{T}$ & $14$ dBm & $R$ & $\left[0;10000\right]$ m\tabularnewline
\hline 
$G_{R,max}$ & $3.2$ dBi & $H$ & $\left[5;120\right]$ m\tabularnewline
\hline 
$BW_{z\xi}$ & $100^{o}$ & $h$ (standing) & $1.7$ m\tabularnewline
\hline 
$\chi$ & $-3$ dB & $h$ (lying) & $0$ m\tabularnewline
\hline 
$f$ & $868$ MHz & $\tau_{T}$ & -0.01 dB\tabularnewline
\hline 
\end{tabular}
\end{table}

\begin{figure}
\begin{centering}
\includegraphics[width=5cm]{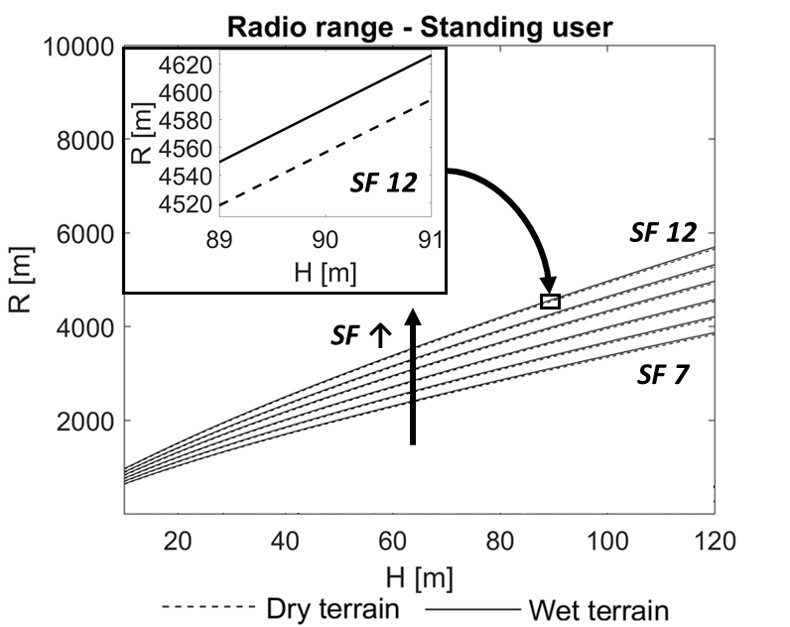}(a)
\par\end{centering}
\begin{centering}
\includegraphics[width=4cm,height=3.5cm]{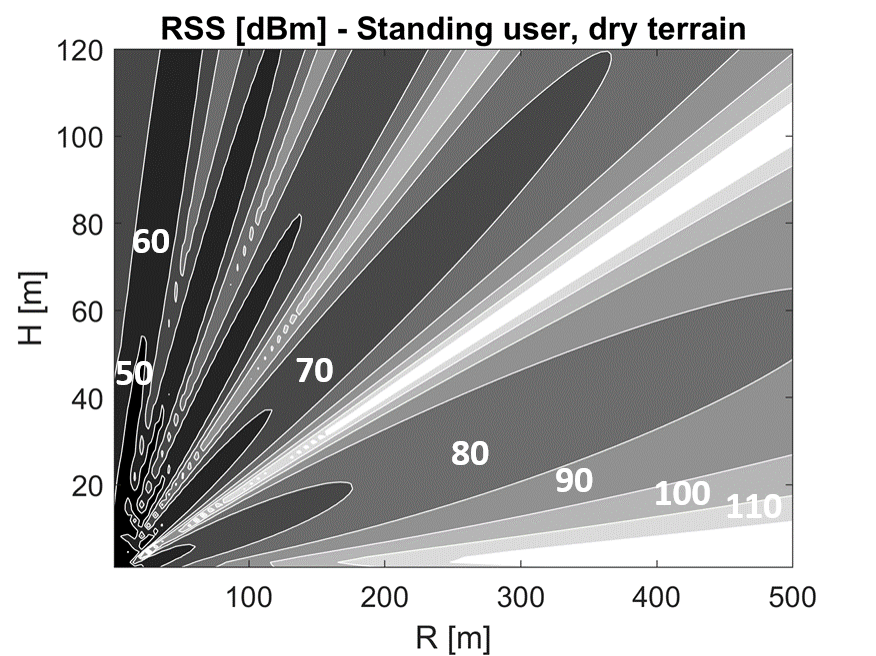}(b)\includegraphics[width=4cm,height=3.5cm]{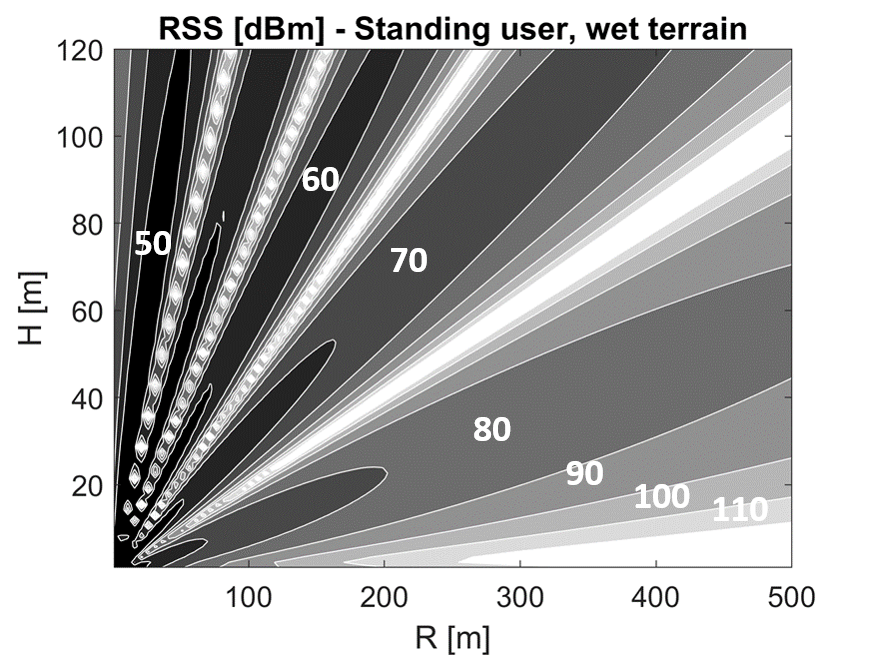}(c)
\par\end{centering}
\centering{}\caption{Theoretical coverage maps of the LoRa transmitting radio-helmet and
the UAV receiver in the standing user case for different flying heights
\emph{H} and radial distances \emph{R}. (a) Maximum communication
distances versus the UAV height when varying the \emph{SF} value in
case of dry and wet terrain. RSS for $R\protect\leq500$ m in case
of (b) dry terrain and (c) wet terrain. \label{fig:standdry}}
\end{figure}

\begin{figure}
\begin{centering}
\includegraphics[width=7.5cm]{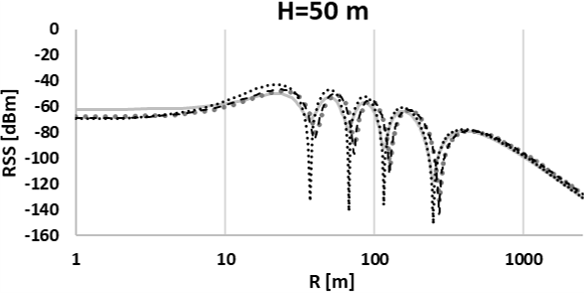}(a)
\par\end{centering}
\smallskip{}

\begin{centering}
\includegraphics[width=7.5cm]{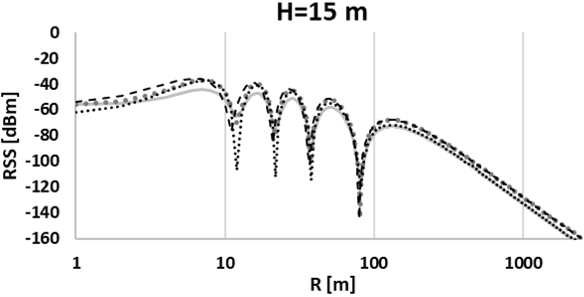}(b)
\par\end{centering}
\begin{centering}
\includegraphics[width=8cm]{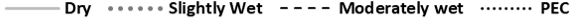}
\par\end{centering}
\caption{\textcolor{black}{Numerically evaluated RSS for $1\ \text{m}\protect\leq R\protect\leq2500\ \textnormal{m}$
and different terrain's wetness. The assumed intermediate permittivities
are $\varepsilon_{0}\left(15-0.4j\right)$ for a slightly wet terrain
and $\varepsilon_{0}\left(30-0.4j\right)$ for a moderately wet terrain
\cite{dry terrain}. Two different flying altitudes are considered
(a) $H=50$ m and (b) $H=15$ m.}\textcolor{blue}{\label{fig:Wetness_Variable}}}
\end{figure}

Although both the lying-user and the standing-user links are in LoS
conditions, the evaluated RSS is rather different. Firstly, by considering
the standing-user case, Fig. \ref{fig:standdry}(a) shows the maximum
estimated communication distances versus \emph{H} for dry or wet terrains.
The link range is mostly unaffected by the terrain's wetness, which
instead impacts the interference fringes, as shown next. The achievable
$R_{max}$ spans from about $3.8$ km to $6.5$ km, depending on the\emph{
SF}. For instance, by increasing $SF$ from $7$ to $12$, the predicted
$R_{max}$ for $H=50$ m lengthens from $2$ km to $3$ km. Focused
coverage maps for \emph{R} $\leq500$ m, wherein the interference
pattern is highly variable, are depicted in Fig. \ref{fig:standdry}(b,c).
The RSS is not monotonically related to the Tx-Rx distance, even in
the ideal free-space case. The nulls in the RSS patterns are sharp
due to the strong interference, especially for short horizontal distances.
Regarding the ground's condition, the wetter the terrain is, the stronger
the reflected field is and, consequently, the more relevant the fluctuations
are. An example of RSS profiles due to different terrain's wetness
is shown in Fig. \ref{fig:Wetness_Variable}. When comparing Fig.
\ref{fig:Wetness_Variable}(a) with Fig. \ref{fig:Wetness_Variable}(b),
it is evident that the fluctuations can be restricted to shorter distances
$R$ by lowering $H$. In particular, for $H=15$ m, the interference
fringes are restricted to $R<100$ m.

\begin{figure}
\begin{centering}
\includegraphics[width=4cm,height=3.5cm]{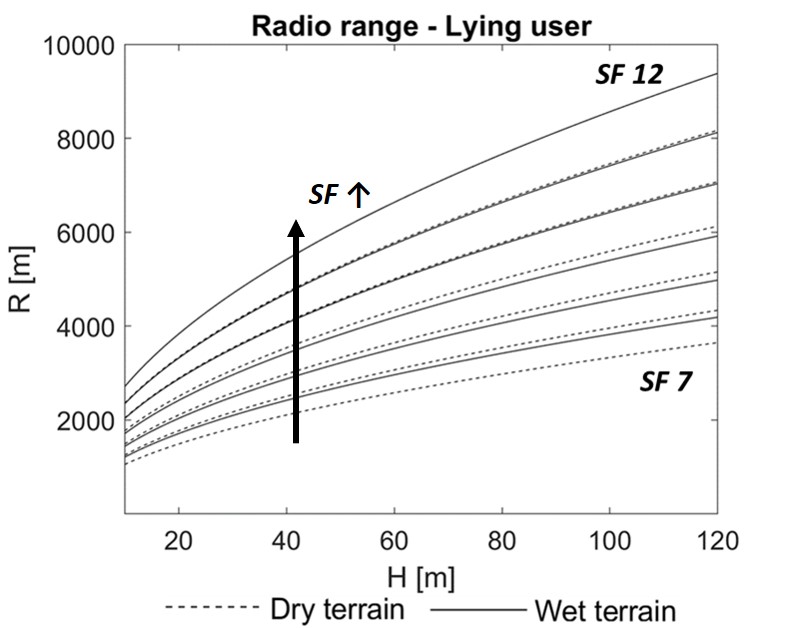}(a)\includegraphics[width=4cm,height=3.5cm]{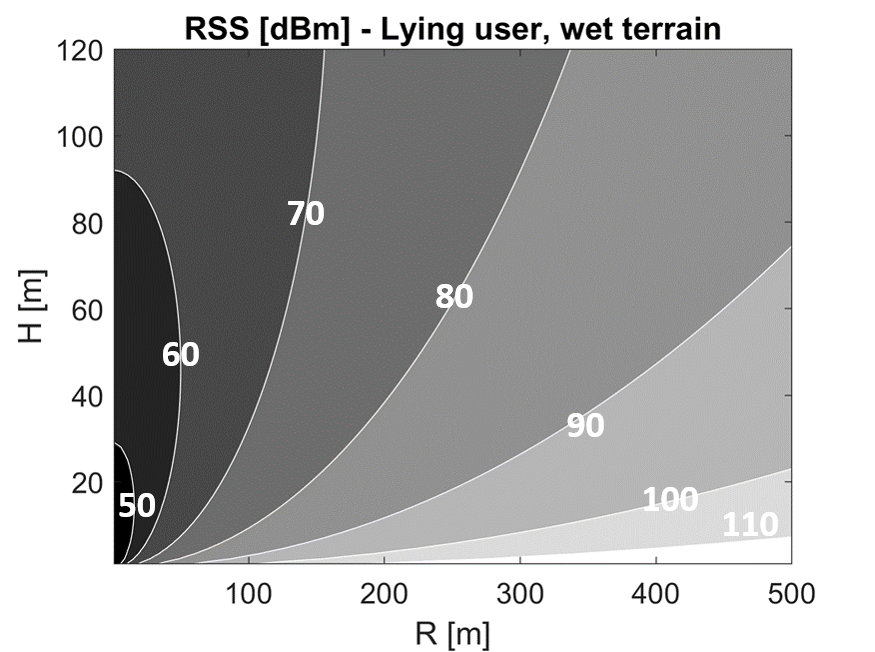}(b)
\par\end{centering}
\caption{Theoretical coverage maps of the LoRa transmitting radio-helmet and
the UAV receiver in the lying user case. (a) Maximum communication
distances versus the UAV height when varying the \emph{SF} value and
(b) RSS for $R\protect\leq500$ m (wet terrain). \label{fig:Coverage-map-fallen}}
\end{figure}

In the case of a lying user, the multi-path is absent so that the
RSS decreases monotonically with the Tx-Rx distance (Fig. \ref{fig:Coverage-map-fallen}).
Unlike the previous case, the wetness of the terrain has a significant
impact on the maximum range, especially for high \emph{SF} values.
Indeed, if the wearer is lying on the ground, the on-helmet antenna's
gain improves with the terrain wetness of $2$ dB (as in Tab. \ref{tab:The-lower-bound-gains}),
at most, leading to a $21$\% higher radio range. The maximum communication
range is up to $9.5$ km if the terrain is wet, and hence it is more
extended than the corresponding distance in the case of a standing
user. 

\section{Experimentation \label{sec:Measured-Signal-Strength}}

The above numerical achievements are corroborated by preliminary experimentation
with a UAV equipped with a LoRa receiver and a typical mountain helmet
embedding a transmitter to reproduce the simulated scenarios. Both
vertical and horizontal flights are performed to verify the model's
prediction.

The LoRa transmitting and receiving modules are Pycom LoPy-4 programmable
boards (embedding the LoRa SX1276 transceiver; $P_{T}=14$ dBm, bandwidth
of $125$ kHz, carrier frequency $868$ MHz, coding rate $4/5$ \cite{IoTJ}).
The spreading factor is set to $SF=12$ to maximize the receiver sensitivity
and characterize the link as far as possible.

\begin{figure}
\begin{centering}
\includegraphics[width=2.25cm,height=2.5cm]{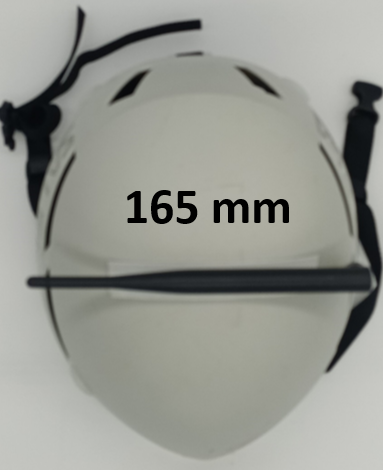}(a)\includegraphics[width=5.75cm,height=2.5cm]{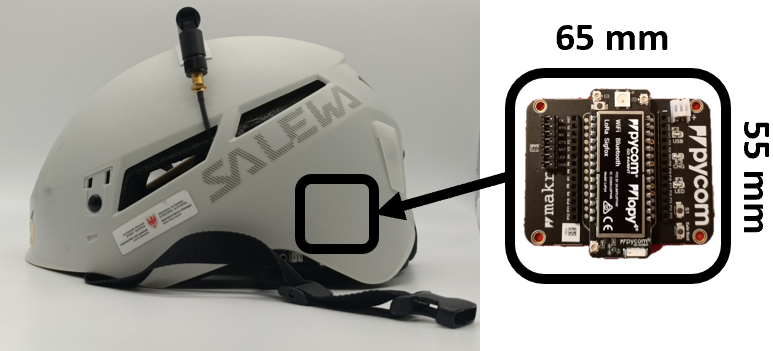}(b)
\par\end{centering}
\caption{Radio-helmet prototype hosting a coronal dipole, (a) top view and
(b) side view with the indication of the LoPy-4 board installed in
the interior of the helmet.\label{fig:Prototype-of-the}}
\end{figure}

The transmitting antenna is a commercial dipole (by Pycom company
\cite{LoRaDipole}) fixed on a Vayu 2.0 helmet for mountaineering
along the coronal plane (Fig. \ref{fig:Prototype-of-the}). The helmet
is worn by a 1.7 m tall volunteer as in the simulations. 

\begin{figure}
\begin{centering}
\textbf{Measured $\Gamma_{\textnormal{T}}$ - Commercial coronal dipole}
\par\end{centering}
\begin{centering}
\includegraphics[width=9cm]{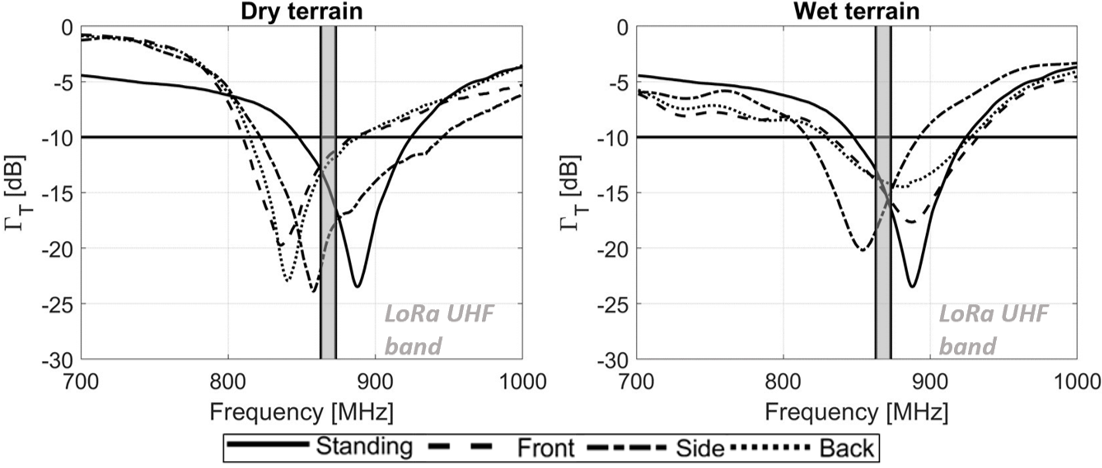}
\par\end{centering}
\caption{The measured reflection coefficient of the coronal dipole installed
over the helmet when the wearer is standing and then\textbf{ }lying
on dry or wet terrain in the three postures. \label{fig:Measured--S11}}
\end{figure}

The input parameter of the helmet-mounted dipole is measured by a
portable Vector Network Analyzer (MS2024A by Anritsu) when the volunteer
is standing and lying over dry and wet terrains. Overall, despite
some frequency shift between standing and fallen positions w.r.t.
the nominal LoRa band, as predicted by the simulations (Fig. \ref{fig:Reflecion-coefficient-S11}),
the obtained reflection coefficient (Fig. \ref{fig:Measured--S11})
is always $\Gamma_{T}<11$ dB at the useful frequencies, and it is
well comparable with the simulated $\tau_{T}$ value in Table \ref{tab:Parameters-considered-in}.
The maximum observed downward frequency shift is about $55$ MHz over
all the conditions, so the helmet antenna must thus ensure a frequency
bandwidth of at least $65$ MHz to cover the whole UHF LoRa band regardless
of the wearer posture. 

\begin{figure}
\begin{centering}
\includegraphics[width=7.5cm]{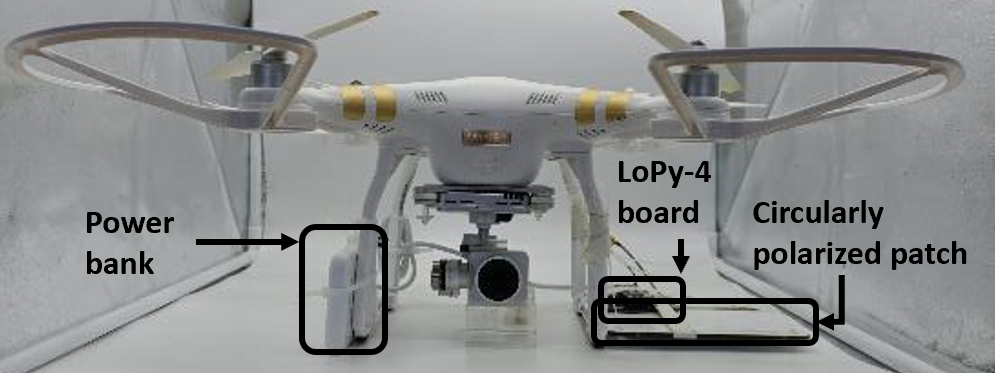}
\par\end{centering}
\caption{DJI Phantom 3 Pro UAV equipped with a LoRa receiver. The LoRa receiver
is powered by a power bank and is composed of a LoPy-4 board and a
Keonn Advantenna-p11 CP patch. \label{fig:receving_UAV}}
\end{figure}

The UAV is a DJI Phantom 3 Pro hosting a $137$ mm $\times137$ mm
CP patch (Advantenna-p11 Keonn as before). The battery pack and the
LoRa receiver are fixed at the landing gear of the UAV (Fig. \ref{fig:receving_UAV}).

\begin{figure}
\begin{centering}
\includegraphics[width=7.5cm]{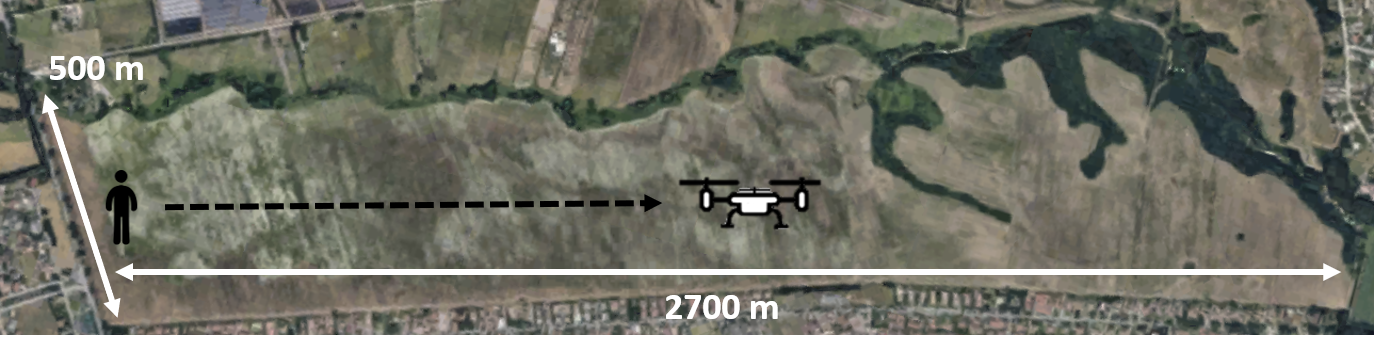}(a)
\par\end{centering}
\begin{centering}
\includegraphics[width=7.5cm]{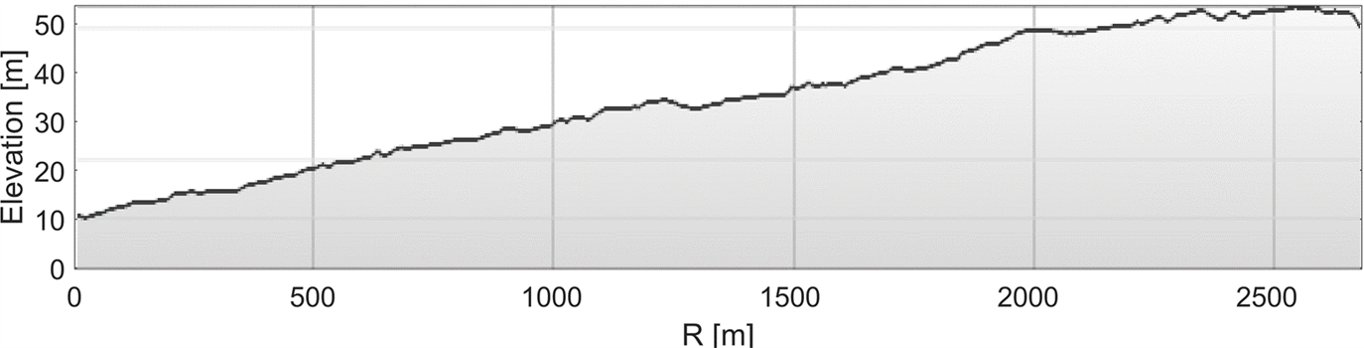}(b)
\par\end{centering}
\begin{centering}
\includegraphics[width=3.5cm]{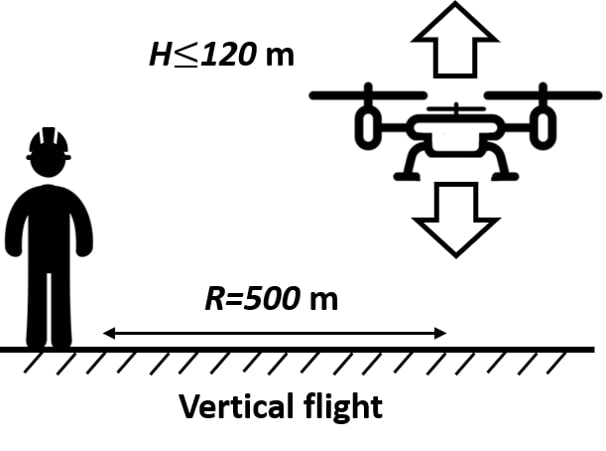}(c)\includegraphics[width=3.5cm]{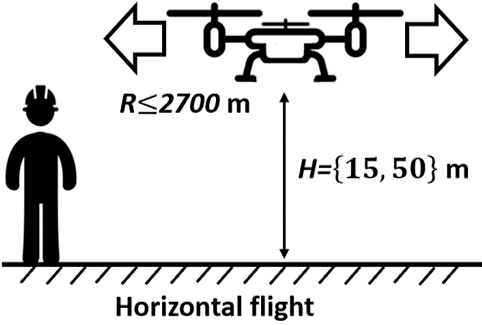}(d)
\par\end{centering}
\caption{(a) Satellite view of the location for the experimental helmet-UAV
LoRa link measurem\textcolor{black}{ent, (b) elevation profile of
the field, (c}) sketch of the vertical flights, and (d) of the horizontal
flights. \label{fig:Field-selected-for-Flights}}
\end{figure}
The RSS measurements are performed during a period of consecutive
sunny days over an uncultivated field near Colle Romito\footnote{GPS coordinates (41.551526, 12.567001).}
{[}Ardea, Lazio, Italy; Fig. \ref{fig:Field-selected-for-Flights}(a){]}.
The field is flat {[}Fig. \ref{fig:Field-selected-for-Flights}(b){]}
and covered by bushes. Thus, the test field can be considered dry
ground. In the experiments, the UAV is driven at $1$ m/s approximate
speed along both horizontal trajectories at fixed altitudes and vertical
trajectories at a fixed radial distance from the user. The RSS in
the dBm scale is estimated from the RSSI (received signal strength
indicator) and from the SNR (signal-to-noise ratio) that are returned
by the receiver, as \cite{IoTJ}

\begin{equation}
P_{R}=RSSI-10\log_{10}\left(1+\frac{1}{SNR}\right)+c_{0},\label{eq:RSS_From_RSSI}
\end{equation}
where $P_{R}$ is in dBm scale, the RSSI is in dB scale, the SNR is
in linear scale, and $c_{0}$ (dBm scale) is a constant parameter
obtained from a single-point calibration \cite{LoRaSensitivities}
through the Anritsu MS2711A spectrum analyzer.

In the first test, the UAV takes off at a distance $R=500$ m from
the volunteer and then flies up vertically up to $H=120$ m {[}Fig.
\ref{fig:Field-selected-for-Flights}(c){]}. In the second test, the
UAV flies horizontally at two fixed altitudes $H=\{$$50$ m, $15$
m\} for distances $1$ m $\leq R\leq2700$ m {[}Fig. \ref{fig:Field-selected-for-Flights}(d){]}.
During the flights, $1921$ data packets are received overall.

\begin{figure}
\begin{centering}
\textbf{\large{}Vertical flights}{\large\par}
\par\end{centering}
\smallskip{}

\begin{centering}
\includegraphics[width=4cm,height=3cm]{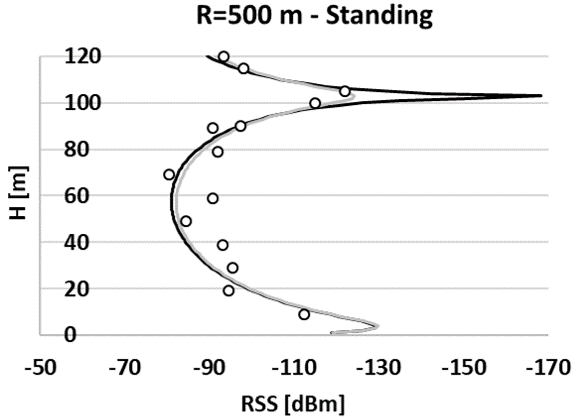}(a)\includegraphics[width=4cm,height=3cm]{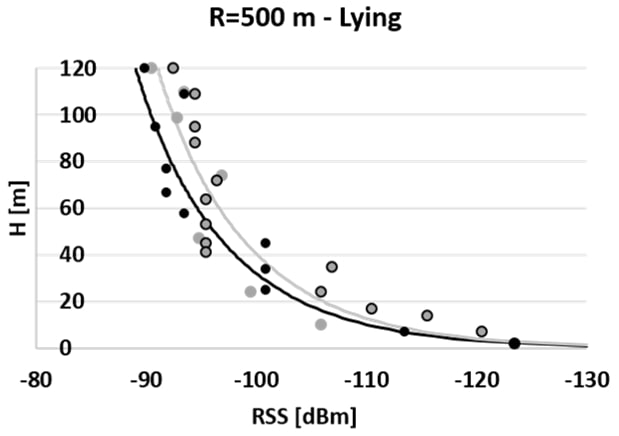}(b)
\par\end{centering}
\begin{centering}
\includegraphics[width=5.5cm]{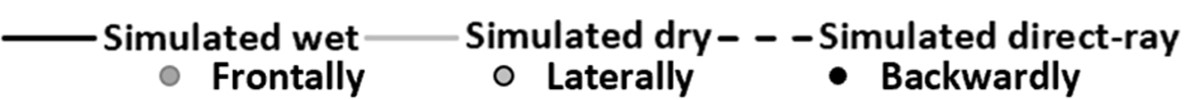}
\par\end{centering}
\caption{RSS measured by the on-UAV \textcolor{black}{LoRa ($SF=12$) }receiver
during the vertical flights at fixed ground distance \emph{R} $=500$
m from the transmitter. Measurements for a (a) standing user and (b)
lying user.\label{fig:RSS-Vertical-Flights}}
\end{figure}

\begin{figure}
\begin{centering}
\textbf{\large{}Horizontal flights}{\large\par}
\par\end{centering}
\smallskip{}

\begin{centering}
\includegraphics[width=4cm]{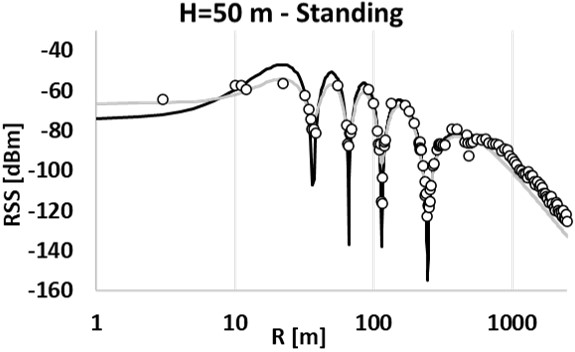}(a)\includegraphics[width=4cm]{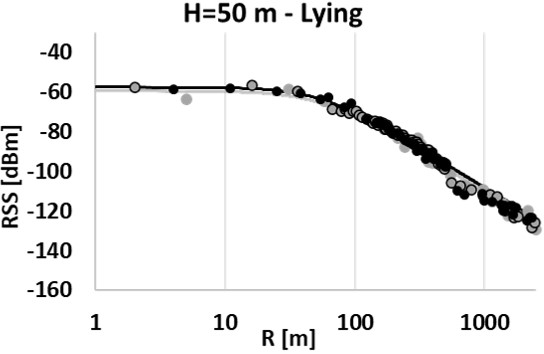}(b)
\par\end{centering}
\begin{centering}
\includegraphics[width=4cm]{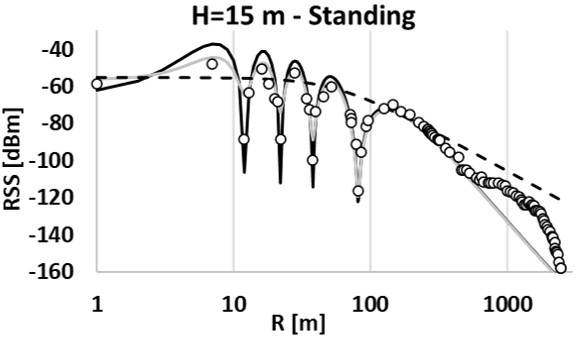}(c)\includegraphics[width=4cm]{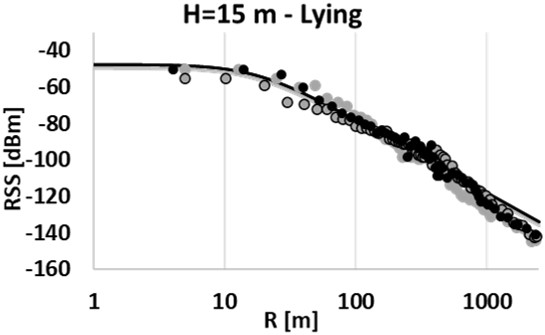}(d)
\par\end{centering}
\begin{centering}
\includegraphics[width=7cm]{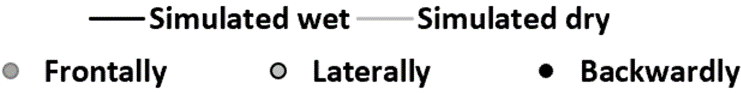}
\par\end{centering}
\caption{RSS measured by the on-UAV LoRa ($SF=12$) receiver during the horizontal
flights at different fixed flying heights \emph{H} and for different
user positions: (a) \emph{H} $=50$ m, standing user, (b) \emph{H}
$=50$ m, lying user, (c) \emph{H} $=15$ m, standing user, and (d)
\emph{H} $=15$ m, lying user. \label{fig:RSS-seen-Horizontal}}
\end{figure}

The measured RSS profiles are reported in Fig. \ref{fig:RSS-Vertical-Flights}
and Fig. \ref{fig:RSS-seen-Horizontal} and compared with the simulated
ones for vertical and horizontal flights, respectively. Interestingly,
the RSS is rather insensitive to the lying user's particular position,
and the modeling through a numerically evaluated equivalent gain is
proven to be effective.

Overall, in spite of some differences, the measurements corroborate
the phenomena derived by the numerical model in section \ref{sec:Link-evaluation}
(Fig. \ref{fig:standdry} and Fig. \ref{fig:Coverage-map-fallen}),
even in the presence of the significant multi-path in the standing
configuration. In detail, simulations overestimate the RSS of just
$2$ dB on the vertical and horizontal paths shown in Fig. \ref{fig:RSS-Vertical-Flights}
and Fig. \ref{fig:RSS-seen-Horizontal}(a,b,d). Regarding Fig. \ref{fig:RSS-seen-Horizontal}(c),
there are some differences with the simulations at the end of the
profile, where the model underestimates the RSS. The measured profile
follows the flat-earth two-ray model for $R\leq700$ m but, afterwards,
it shows an intermediate behavior between the two-ray and the single-ray
propagation. In this condition, the incident rays intercept the ground
at grazing angles ($\varphi\leq1.4^{\textnormal{o}}$) so that slight
surface irregularities are no longer negligible \cite{CoverageMaps}.
Moreover, the combined effect of the absorption by the low vegetation,
scattering, and the earth elevation {[}Fig. \ref{fig:Field-selected-for-Flights}(b){]}
can attenuate further the reflected field. The result is a weaker
interference and, consequently, a lengthening of the link range, benefiting
the target identification from long distances.

The Tx-Rx distance slightly affects also the time delay between the
transmission and the reception of the packets. In the performed flights,
the measured delay ranges between 820 ms (at the shortest helmet-UAV
mutual distance) up to 1022 ms (upper bound distance) and is comparable
with the values reported in \cite{Widianto18}. By considering the
typical packet transmission rate (roughly 1 packet every 3 s), the
above delays will not cause any loss of the received packet sequence.
Accordingly, the observed packet delivery ratio (PDR) in the experiments
is of the order of 95\% and nearly constant in all the cases. In more
complex environments with non-flat grounds, the soft degradation of
the LoRa signals will lead to reduced radio ranges with the corresponding
decrease of the PDR \cite{LoRarange,IoTJ}. Concerning the implications
on localization, the time delay could slightly raise the localization
errors, especially if the target is moving. In SaR operations, however,
the transmitter is mostly stationary since the wearer could be fallen
or at most slowly moving within a confined area. The corresponding
error in estimating the position of the target is therefore expected
to be modest. Furthermore, the UAV will generally get closer and closer
to the target during SaR procedures so that the time delay reduces,
the PDR increases, and the estimation of the target position is progressively
refined.

It is worth specifying that, due to hardware constraints and law regulations,
the above experimental tests did not allow us to appreciate the complete
communication range corresponding to the last received packet. For
this purpose, the UAV should have flown up to a much longer horizontal
distance ($R>2700$ m), thus exiting from the test area and overflying
various obstacles (houses, roads, woods). Anyway, in past experiments
\cite{Splitech}, we placed a LoRa radio over an $85$-m-high overpass
while the other radio was kept inside a car that was driven away from
the overpass. This arrangement, comparable with the helmet-UAV system,
achieved communications up to $8.4$ km, in reasonable agreement with
the numerical estimations.

\section{Discussion and conclusion\label{sec:Conclusions}}

The helmet-to-UAV communication over flat lands based on the LoRa
protocol has been numerically characterized and corroborated by a
preliminary experimental campaign. The UAV is theoretically capable
of collecting the helmet signal up to $5$ km in a flat scenario even
in the worst case (standing user, perfectly flat and wet terrain),
and up to $9.5$ km in case of wet terrain and lying user when the
UAV flies at an altitude of $120$ m. The RSS in the case of a lying
user is robust concerning the position of the helmet on the ground,
and it monotonically decreases with the Tx-Rx distance. Instead, the
RSS with a standing user (e.g., a lost hiker wandering on a snowy
land) experiences fluctuations that are sharper the wetter the terrain,
due to the ground-bounce multi-path, especially near to the helmet's
wearer.

The basic model extension to a more complex radio propagation environment,
including non-LoS links, can be quickly drawn if an equivalent log-distance
PL model is available. For example, the authors recently modeled the
LoRa propagation in snowy mountain fields, even when the transmitter
is buried \cite{IoTJ}. Burying the transmitter under $1$ m of snow
adds an entry loss of about $60$ dB, which can be accounted for in
the transmitter's equivalent gain. Instead, the retrieved log-distance
parameters for a transmitter placed on the snow were $n=3.17$ and
$PL(d_{0}=1\ \textnormal{m})=56.7$ dB, and these values are still
approximately valid in the case of a lying user in a snowy plain.
Hence, the path loss model and the eventual path-gain factor must
be inserted appropriately in the proposed link budget.

It is worth considering that the communication could be more challenging
or even forbidden in harsher conditions that excessively reduce the
equivalent antenna gain, as, for instance, when the helmet is buried
under more than 1 m of wet snow or deeply submerged by water (e.g.,
fallen into a river). In other cases, particular configurations of
obstacles may highly raise the multi-path and the shadowing, as we
observed when the target was on the bottom of a canyon while the UAV
was flying outside it \cite{MejiaAguilar21In}.

Finally, considering the model's application to SaR operations, the
lying user case is the most interesting because the user could be
unconscious and even partially covered and, consequently, not detectable
by a camera. The helmet wearer can be rescued through classic range-based
localization algorithms like those mentioned in section \ref{sec:Introduction}.
Instead, in the case of a standing user, the interference fringes
make applying the above localization methods more challenging \cite{Thompson12}.
Spatial fluctuations can be mitigated by lowering the UAV's flying
altitude. For instance, the RSS oscillations are confined between
$100$ m far from the target for an altitude $H=15$ m. The fluctuations
can also be reduced by placing the wearable antenna near the ground
(e.g., on a boot). In any case, thanks to the local confinement of
RSS fluctuations, the UAV can roughly estimate the target position
from longer distances. Then, the UAV can approach the estimated location
and, when the signal begins fluctuating, it could use the support
of optical or thermal cameras to reach the user, who is standing and,
likely, conscious. Hence, while the high-RSS fringes are produced
by the reflection from the flat ground when the user is standing,
the eventual low-RSS fringes are generated by the multi-path from
non-flat terrains and will be observed for longer distances. Furthermore,
the dynamic refinement of the position estimation while approaching
the target allows also to contrast the packet's loss caused by the
harsh environment. Regardless of the target's posture, the UAV could
move along non-linear trajectories to speed up the search, as proposed
in \cite{Buffi2019,Shahidian16}, by dynamically accounting for the
signal variations and the topology of the area. A swarm of multiple
coordinated UAVs \cite{Sallouha2018,Chen2020} could also be deployed
to provide data from different angles and distances to mitigate the
multi-path effect.


\begin{thebibliography}{10}
\bibitem{SaR_Fire_Flood} \textquotedblleft Emergency Support Function
\#9: SEARCH AND RESCUE'', N.H. Dept. of Fish and Game, Concord, New
Hampshire State, USA, Rep. no ESF\#9, 2006. Accessed: Feb. 12, 2021.
{[}Online{]}. Available: https://prd.blogs.nh.gov/dos/hsem/wp-content/uploads/2015/03/ESF-9-Search-Rescue.pdf

\bibitem{SaR_earthquakes} W.-T. Chiu \emph{et al.}, ``A survey of
international urban search-and-rescue teams following the Ji Ji earthquake'',
\emph{Disasters}, vol. 26, no. 1, pp. 85-94, Dec. 2002.

\bibitem{Accidentology} B. Soule, B. Lefevre, E. Boutroy, V. Reynier,
F. Roux, and J. Corneloup, \emph{Accidentology of Mountain Sports:
Situation, Review \& Diagnosis}, Fondation Petzl, Crolles, France,
2014.

\bibitem{ARVA} \emph{Avalanche Beacons operating at 457 kHz; Transmitter-receiver
systems, }ETSI EN 300 7181 V2.1.1, European Telecommunications Standards
Institute, 2018. Accessed on: Feb. 12, 2021. {[}Online{]}. Available:
\ https://www.etsi.org/deliver/etsi\_en/300700\_300799/30071801/02.01.00
\_20/en\_30071801v020100a.pdf

\bibitem{RECCO}\emph{ R9 Detector User Guide, }RECCO Company, Liding,
Sweden, 2009. Accessed: Feb. 12, 2021. {[}Online{]}. Available: https://usermanual.wiki
/RECCO/A-

\bibitem{datarate} J. Bardyn, T. Melly, O. Seller and N. Sornin,
``IoT: the era of LPWAN is starting now,'' in\emph{ Proc. 42nd Eur.
Solid-State Circuits Conf}., Lausanne, France, 2016, pp. 25-30.

\bibitem{LoRarange} J. Petajajarvi, K. Mikhaylov, A. Roivainen, T.
Hanninen and M. Pettissalo, ``On the coverage of LPWANs: range evaluation
and channel attenuation model for LoRa technology,'' in \emph{Proc.
14th Int. Conf. ITS Commun.}, Copenhagen, Denmark, 2015, pp. 55-59.

\bibitem{patienttracking} A. T. Nugraha, R. Wibowo, M. Suryanegara
and N. Hayati, ``An IoT-LoRa system for tracking a patient with a
mental disorder: correlation between battery capacity and speed of
movement,'' in \emph{Proc. 7th Int. Conf. Compute. Commun. Eng.},
Kuala Lumpur, Malaysia, 2018, pp. 198-201.

\bibitem{LoRaHealth} J. Petajajarvi, K. Mikhaylov, M. Hemeleinen
and J. Iinatti, ``Evaluation of LoRa LPWAN technology for remote
health and wellbeing monitoring,\emph{'' }in \emph{10th Int. Symp.
Med. Inf. Commun. Technol.,} Worcester, MA, USA, 2016, pp. 1-5.

\bibitem{Shi2021} L. Shi, H. Xu, W. Ji, B. Zhang, X. Sun and J. Li,
``Real-time human activity recognition system based on capsule and
LoRa,'' in \emph{IEEE Sensors J.}, vol. 21, no. 1, pp. 667-677, Jan.
2021.

\bibitem{Yuan18} Z. Yuan, J. Jin, L. Sun, K. Chin and G. Muntean,
``Ultra-reliable IoT communications with UAVs: A swarm use case,''
in \emph{IEEE Commun. Mag}., vol. 56, no. 12, pp. 90-96, Dec. 2018.

\bibitem{Catherwood2020} P. A. Catherwood, M. Little, D. Finlay and
J. McLaughlin Obe, \textquotedbl Recovery of incapacitated commercial
delivery drones using LPWAN technology,\textquotedbl{} in\emph{ IEEE
Intell. Transp. Syst. Mag}., vol. 12, no. 2, pp. 6-19, Apr. 2020.

\bibitem{Splitech} G. M. Bianco, A. Mejia-Aguilar and G. Marrocco,
``Performance evaluation of LoRa LPWAN technology for mountain search
and rescue,'' in \emph{Proc. 5th Int. Conf. Smart Sustain. Technol}.,
Split, Croatia, 2020, pp. 1-4.

\bibitem{IoTJ} G. M. Bianco, R. Giuliano, G. Marrocco, F. Mazzenga
and A. Mejia-Aguilar, ``LoRa system for search and rescue: path-loss
models and procedures in mountain scenarios,'' \emph{IEEE Internet
Things J.}, vol. 8, no. 3, pp. 1985-1999, Feb. 2021.

\bibitem{Sisinni2020} E. Sisinni, D. F. Carvalho and P. Ferrari,
``Emergency communication in IoT scenarios by means of a transparent
LoRaWAN enhancement,'' in \emph{IEEE Internet of Things J.}, vol.
7, no. 10, pp. 10684-10694, Oct. 2020.

\bibitem{Gimenez2020} F. Gimenez, C. Zerbini and G. Riva, ``Extending
SMS service coverage in rural areas by using LoRa communication technology,''
(in Spanish), \emph{IEEE Latin Amer. Trans.}, vol. 18, no. 2, pp.
214-222, Feb. 2020.

\bibitem{Tran2020} H. P. Tran, W.-S. Jung, T. Yoon, D.-S. Yoo and
H. Oh, ``A two-hop real-time LoRa protocol for industrial monitoring
and control systems,'' in \emph{IEEE Access}, vol. 8, pp. 126239-126252,
Jul. 2020, Art. no. 9136688.

\bibitem{Lin2020} K. Lin and T. Hao, \textquotedbl Experimental
link quality analysis for LoRa-based wireless underground sensor networks,\textquotedbl{}
in \emph{IEEE Internet Things J}., early access, Dec. 15 , 2020. doi:
10.1109/JIOT.2020.3044647.

\bibitem{Aslam2020} M. S. Aslam\emph{ et al}., \textquotedbl Exploring
multi-hop LoRa for green smart cities,\textquotedbl{} \emph{IEEE Netw.},
vol. 34, no. 2, pp. 225-231, Mar. 2020.

\bibitem{Pham2020} T. V. Pham \emph{et al}., \textquotedbl Proposed
smart university model as a sustainable living lab for university
digital transformation,\textquotedbl{} in\emph{ Proc. 5th Int. Conf.
Green Technol. Sustain. Develop.}, Ho Chi Minh City, Vietnam, Nov.
2020, pp. 472-479.

\bibitem{Lee2011} J. H. Lee and R. M. Buehrer, ``Fundamentals of
received signal strength-based position location,'' in \emph{Handbook
of Position Location: Theory, Practice, and Advances}, 1st ed, Hoboken,
NJ, USA, John Wiley and Sons, 2012, ch.11, pp. 359-395.

\bibitem{Bianco21Multislope}G. M. Bianco, R. Giuliano, F. Mazzenga
and G. Marrocco, \textquotedbl Multi-slope path loss and position
estimation with grid search and experimental results,\textquotedbl{}
\emph{IEEE Trans. Signal Inf. Process. Netw.}, vol. 7, pp. 551-561,
2021

\bibitem{Gholami13} M. R. Gholami, R. M. Vaghefi, and E. G. Stram,
\textquotedblleft RSS-based sensor localization in the presence of
unknown channel parameters,\textquotedblright{} \emph{IEEE Trans.
Signal Process.}, vol. 61, no. 15, pp. 3752--3759, Aug. 2013.

\bibitem{Li2006} X. Li, \textquotedblleft RSS-based location estimation
with unknown pathloss model,\textquotedblright{} \emph{IEEE Trans.
Wireless Commun.}, vol. 5, no. 12, pp. 3626--3633, Dec. 2006.

\bibitem{disasterprevention} Y. Saita, T. Ito, N. Michishita and
H. Morishita, ``Low-frequency inverted-F antenna on hemispherical
ground plane,'' in\emph{ Proc. Int. Symp. Antennas Propag. Conf.},
Kaohsiung, Taiwan, 2014, pp. 183-184.

\bibitem{radome} D. Kitching and F. Lalezari, ``Low profile helmet
mount GPS antenna,'' in \emph{Proc. Conf. Tactical Commun}., Fort
Wayne, IN, USA, 1990, pp. 661-702.

\bibitem{HelmetsSAR} N. Nishiyama, N. Michishita and H. Morishita,
``SAR reduction of helmet antenna composed of folded dipole with
slit-loaded ring,'' in \emph{Proc. Int. Symp. Antennas Propag.},
Hobart, Australia, 2015, pp. 1-2.

\bibitem{HelmetSAR2} N. Nishiyama, N. Michishita and H. Morishita,
``Low-frequency inverted-F antenna on annular ground plane,'' in
\emph{Proc. IEEE MTT-S Int. Microw. Workshop Series RF Wireless Technol.
Biomed. Healthcare Appl.}, Taipei, Taiwan, 2015, pp. 143-144.

\bibitem{LoopHelmet} J.-Y. Park, H.-K. Ryu and J.-M. Woo, ``Helmet
installed antenna using a half-wavelength circular loop antenna,''
in \emph{Proc. IEEE Antennas Propag. Soc. Int. Symp}., Honolulu, HI,
USA, 2007, pp. 4176-4179.

\bibitem{Kachroo2019} A. Kachroo \emph{et al.}, \textquotedbl Unmanned
aerial vehicle-to-wearables (UAV2W) indoor radio propagation channel
measurements and modeling,\textquotedbl{} in \emph{IEEE Access}, vol.
7, pp. 73741-73750, May 2019.

\bibitem{Ameloot21} T. Ameloot, P. Van Torre and H. Rogier, ``LoRa
base-station-to-body communication with SIMO front-to-back diversity'',
\emph{IEEE Trans. Antennas Propag.}, vol. 69, no. 1, pp. 397-405,
Jan. 2021.

\bibitem{Dumanli17} S. Dumanli, L. Sayer, E. Mellios, X. Fafoutis
and G. S. Hilton, ``Off-body antenna wireless performance evaluation
in a residential environment'', \emph{IEEE Trans. Antennas Propag.},
vol. 65, no. 11, pp. 6076-6084, Nov. 2017.

\bibitem{Olasupo2019} T. O. Olasupo, \textquotedblleft Wireless communication
modeling for the deployment of tiny IoT devices in rocky and mountainous
environments,\textquotedblright{}\emph{ IEEE Sensors Lett.}, vol.
3, no. 7, pp. 1--4, Jul. 2019

\bibitem{Simunek13} M. Simunek, F. P. Fontan and P. Pechac, ``The
UAV low elevation propagation channel in urban areas: statistical
analysis and time-series generator'', \emph{IEEE Trans. Antennas
Propag.}, vol. 61, no. 7, pp. 3850-3858, Jul. 2013.

\bibitem{Cui2019} Z. Cui \emph{et al}., ``Low-altitude UAV air-ground
propagation channel measurement and analysis in a suburban environment
at 3.9 GHz,'' \emph{IET Microw., Antennas Propag.}, vol. 13, no.
9, pp. 1503-1508, Jul. 2019.

\bibitem{Safwat20} N. E.-D. Safwat, F. Newagy and I. M. Hafez, ``Air-to-ground
channel model for UAVs in dense urban environments,'' \emph{IET Microw.,
Antennas Propag.}, vol. 14, no. 6, pp. 1016-1021, Apr. 2020.

\bibitem{LoRaBand} K. Mikhaylov, J. Petajajarvi and T. Hanninen,
``Analysis of capacity and scalability of the LoRa low power wide
area network technology,'' in\emph{ 22nd Eur. Wireless Conf}., Oulu,
Finland, 2016, pp. 1-6.

\bibitem{dry terrain} \emph{Electrical characteristics of the surface
of the Earth}, ITU-R P.527-4, International Telecommunication Union
Radiocommunication Sector, 2017. Accessed: Feb. 12, 2021. {[}Online{]}.
Available: https://www.itu.int/rec/R-REC-P.527-4-201706-I/en

\bibitem{UAV_GtoA_Propagation} A. A. Khuwaja, Y. Chen, N. Zhao, M.
Alouini and P. Dobbins, \textquotedblleft A survey of channel modeling
for UAV communications,\textquotedblright{} \emph{IEEE Commun. Surv.
Tut.}, vol. 20, no. 4, pp. 2804-2821, doi: 10.1109/COMST.2018.2856587.

\bibitem{LoRaSensitivities} \emph{SX1272/73 Datasheet}, Semtech Corporation,
Camarillo, CA, USA, 2017. Accessed: Feb. 12, 2021. {[}Online{]}. Available:
https://www.mouser.com/datasheet/2/761/sx1272-1277619.pdf

\bibitem{TransmissionParameersLoRa} M. Bor and U. Roedig, ``LoRa
transmission parameter selection,'' in \emph{Proc. 13th Int. Conf.
Distrib. Comput. Sensor Syst.}, Ottawa, Canada, 2017, pp. 27-34.

\bibitem{CoverageMaps} R. E. Collin, ``Radio-Wave Propagation'',
in \emph{Antennas and Radiowave Propagation}, New York, NY, USA, Mc-Graw
Hill, 1985, ch. 6, sec. 1, pp. 339-349.

\bibitem{UAVFootprint}G. Casati \emph{et al.}, ``The interrogation
footprint of RFID-UAV: electromagnetic modeling and experimentations,''
\emph{IEEE J. Radio Freq. Identification}, vol. 1, no. 2, pp. 155-162,
June 2017.

\bibitem{ellipsoid} G. Marrocco, E. Di Giampaolo and R. Aliberti,
``Estimation of UHF RFID reading regions in real environments,''\emph{
IEEE Antennas Propag. Mag}., vol. 51, no. 6, pp. 44-57, Dec. 2009.

\bibitem{headSAR} \emph{IEEE Recommended Practice for Determining
the Peak Spatial-Average Specific Absorption Rate (SAR) in the Human
Head from Wireless Communications Devices: Measurement Techniques},
IEEE 1528-2013, IEEE Standards Association, Sept. 2013. {[}Online{]}.
Available: https://standards.ieee.org/standard/1528-2013.html

\bibitem{foamcostant} Y. H. Kim, S. W. Chan, J. Ahn and S. H. Cho,
``Studies of the variation in the dielectric constant and unique
behaviors with changes in the foaming ratio of the microcellular foaming
process'', in \emph{Polym.-Plastics Technol. Eng}., vol. 50, no.
8, pp. 762-767, June 2011.

\bibitem{UAV_open_category} ``Safe operation of drones in Europe'',
European Union Aviation Safety Agency, April 2018. Accessed: Nov.
5, 2020. {[}Online{]}. Available: https://www.easa.europa.eu/sites/default/files/dfu/217603\_EASA\_
DRONES\_LEAFLET\%20\%28002\%29\_final.pdf

\bibitem{EIRP_limit_LoRa} M. Saelens, J. Hoebeke, A. Shahid and E.
De Poorter, ``Impact of EU duty cycle and transmission power limitations
for sub-GHz LPWAN SRDs: an overview and future challenges,'' in\emph{
EURASIP J. Wireless Commun. Netw.}, vol. 2019, Sept. 2019, Art. no.
219, doi: 10.1186/s13638-019-1502-5.

\bibitem{LoRaDipole} \emph{LoRa (868MHz/915MHz) \& Sigfox Antenna
Kit}, Pycom Company, Guildford, U.K., 2017. Accessed on: Feb. 12,
2021. {[}Online{]}. Available: https://pycom.io/product/lora-868mhz-915mhz-sigfox-antenna-kit/

\bibitem{Widianto18} E. D. Widianto, M. S. M. Pakpahan, A. A. Faizal
and R. Septiana, \textquotedbl LoRa QoS performance analysis on various
spreading factor in Indonesia,\textquotedbl{} \emph{Int. Symp. Electron.
Smart Devices}, 23-24 Oct. 2018, Bandung, Indonesia, pp. 1-5.

\bibitem{MejiaAguilar21In} A. Mejia-Aguilar, G. M. Bianco, G. Marrocco,
A. Voegele, M. van Veelen and G. Strapazzon, \textquotedblleft In-situ
and proximal sensing techniques for monitoring natural hazards to
mitigate risk in tourism activities: A case study in the Geoparc Bletterbach,
Italy\textquotedblright , \emph{IEEE Int. Geosci. Remote Sens. Symp.},
Brussels, Belgium, 11-16 July 2021.

\bibitem{Thompson12}R. J. R. Thompson, E. Cetin, and A. G. Dempster,
\textquotedblleft Unknown source localization using RSS in open areas
in the presence of ground reflections,\textquotedblright{} in \emph{Proc.
2012 IEEE/ION Position, Location Navigation Symp.}, Myrtle Beach,
SC, USA, Apr. 2012, pp. 1018--1027.

\bibitem{Buffi2019} A. Buffi, A. Motroni, P. Nepa, B. Tellini and
R. Cioni, \textquotedbl A SAR-based measurement method for passive-tag
positioning with a flying UHF-RFID reader,\textquotedbl{} \emph{IEEE
Trans. Instrum. Meas}., vol. 68, no. 3, pp. 845-853, Mar. 2019.

\bibitem{Shahidian16} S. A. A. Shahidian and H. Soltanizadeh, ``Optimal
trajectories for two UAVs in localization of multiple RF sources,''\emph{
Trans. Inst. Meas. Control}, vol. 38, no. 8, pp. 908-916, Aug. 2016.

\bibitem{Sallouha2018} H. Sallouha, M. M. Azari, A. Chiumento and
S. Pollin, \textquotedbl Aerial anchors positioning for reliable
RSS-based outdoor localization in urban environments,\textquotedbl\emph{
IEEE Wireless Commun. Lett}., vol. 7, no. 3, pp. 376-379, June 2018.

\bibitem{Chen2020}Y.-J. Chen, D.-K. Chang and C. Zhang, \textquotedbl Autonomous
tracking using a swarm of UAVs: a constrained multi-agent reinforcement
learning approach,\textquotedbl{} \emph{IEEE Trans. Veh. Techno}l.,
vol. 69, no. 11, pp. 13702-13717, Nov. 2020
\end{thebibliography}
\end{document}